\documentclass[reprint,
 amsmath,amssymb,
 aps,pra 
]{revtex4-2}

\usepackage{graphicx}
\usepackage{dcolumn}
\usepackage{bm}
\usepackage[version=4]{mhchem}

\usepackage[normalem]{ulem}

\begin{document}

\title{Charge migration manifests as attosecond solitons in conjugated organic molecules}

\author{Fran\c{c}ois Mauger$^{1}$, 
Aderonke S.\ Folorunso$^{2}$,
Kyle A.\ Hamer$^{1}$,
Cristel Chandre$^{3}$,
Mette B.\ Gaarde$^{1}$,
Kenneth Lopata$^{2,4}$, and
Kenneth J.\ Schafer$^{1}$}
\affiliation{%
$^{1}$Department of Physics and Astronomy, Louisiana State University,
    Baton Rouge, Louisiana 70803, USA \\ 
$^{2}$Department of Chemistry, Louisiana State University,
    Baton Rouge, Louisiana 70803, USA \\
$^{3}$CNRS, Aix Marseille Univ, I2M, Marseille, France \\
$^{4}$Center for Computation and Technology, Louisiana State University, Baton Rouge, Louisiana 70803, USA 
}%

\date{\today}

\begin{abstract}
Charge migration is the electronic response that immediately follows localized ionization or excitation in a molecule, before the nuclei have time to move. It typically unfolds on sub-femtosecond time scales and most often corresponds to dynamics far from equilibrium, involving multi-electron interactions in a complex chemical environment.
While charge migration has been documented experimentally and theoretically in multiple organic and inorganic compounds, the general mechanism that regulates it remains unsettled. 
In this work we use tools from nonlinear dynamics to analyze charge migration that takes place along the backbone of conjugated hydrocarbons, which we simulate using time-dependent density functional theory. 
In this electron-density framework we show that charge migration modes emerge as attosecond solitons and demonstrate the same type of solitary-wave dynamics in both simplified model systems and full three-dimensional molecular simulations.
We show that these attosecond-soliton modes result from a balance between dispersion and nonlinear effects tied to time-dependent multi-electron interactions.
Our soliton-mode mechanism, and the nonlinear tools we use to analyze it, pave the way for understanding migration dynamics in a broad range of organic molecules.
For instance, we demonstrate the opportunities for chemically steering charge migration via molecular functionalization, which can alter both the initially localized electron perturbation and its subsequent time evolution.
\end{abstract}

\maketitle


\section{Introduction}

The movement of electrons and holes in matter regulates many physical and chemical processes such as chemical reactions, photosynthesis and photovoltaics, and charge transfer~\cite{Krausz2009,Calegari2016}. For the electrons, these dynamics can reach down to the Angstrom and attosecond spatio-temporal scales, and are commonly referred to as charge migration (CM)~\cite{Breidbach2005,Lunnemann2008,Smirnova2009,Calegari2014,Kraus2015,Kuleff2016}. At the fastest time scales, CM is understood as the purely electronic-driven dynamics that takes place before nuclei have time to move. It can be the precursor for many of the down-stream processes mentioned above~\cite{Calegari2016,Remacle2006,Lepine2014,Despre2015}, and therefore a means of understanding and ultimately steering them with the goal of charge-directed reactivity~\cite{Weinkauf1997,Remacle1998}. 

The study of molecular CM is a formidable endeavor. 
Experimental studies require coherent probes with attosecond resolution~\cite{Calegari2014,Calegari2016}. Such ultrafast probes have been enabled by the continuous progress in x-ray and table-top infrared sources over the past few decades~\cite{Krausz2009}.
For example, recent experiments have investigated how fast CM couples to the nuclear degrees of freedom, leading to structural changes~\cite{Lara-Astiaso2018,Mansson2021}.
Theoretical investigations of CM necessitate models with multiple interacting electrons~\cite{Vacher2017,Despre2018}, even when nuclear motion is ignored~\cite{Ayuso2017,Bruner2017,Jia2017,Perfetto2018,Folorunso2021}. Furthermore, systematic studies of the underlying mechanisms responsible for regulating CM involve the analysis of systems with a large number of coupled degrees of freedom. 

In this work, we leverage tools from nonlinear dynamics to study (sub-)femtosecond field-free CM unfolding in conjugated organic molecules following the sudden creation of a localized hole in the system, a paradigm for site specific ionization. 
This is motivated by the fact that nonlinear dynamics~\cite{Masoliver2011,RegrAndChaosDyn} has developed general-purpose methods for tackling and understanding the structure of high-dimensional phase spaces.  Indeed, nonlinear dynamical analyses have been instrumental in many areas of atomic, molecular and optical science, including in transition-state theory of chemical reactions~\cite{Miller1998,Bartsch2005,Kawai2007} and strong-field physics~\cite{vandeSand1999,Mauger2009,Mauger2010,Becker2012}.

\begin{figure}
    \centering
    \includegraphics[width=.8\linewidth]{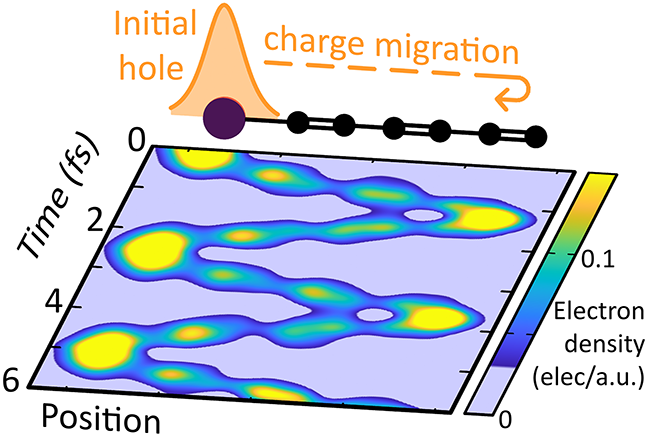}
    \caption{\label{fig:Fig_1}
    Charge migration (CM) manifests as attosecond solitons in conjugated organic molecules:
    (top) Schematic of our investigations of CM in the $\pi$ system of a conjugated carbon chain. We initially put a localized hole at one end of the conjugated system and then (colormap) track the field-free CM electron dynamics it induces in the molecule.
    }
\end{figure}

We show that periodic modes of CM that feature a localized hole that travels back and forth along the molecule's conjugated chain, as sketched in figure~\ref{fig:Fig_1}, emerge as attosecond solitons. These solitary waves represent a balance between dispersion and nonlinear effects that are driven by time-dependent multi-electron interactions.
Previous studies have invoked the beating between a few molecular orbitals as the underlying mechanism that regulates CM~\cite{Remacle2006,Calegari2014,Ayuso2017}. Instead, here we propose a different view of the CM dynamics, directly in the time domain.
Our attosecond-soliton picture provides a generic mechanism for sustained CM motions in organic molecules that does not rely on nonphysical molecular orbitals to describe CM dynamics.
Notably,  we find that the same molecule can support several of these CM modes with periods varying by several hundred attoseconds.
We discuss the implication of our results for future theoretical and experimental CM studies, including the opportunities for chemically steering CM by molecular functionalization, both in creating the initially localized electron hole and for its subsequent time evolution.

The paper is organized as follows:
Section~\ref{sec:CM_dynamics} introduces the theoretical and computational framework we use to study CM dynamics. 
Section~\ref{sec:nonlinear_dynamics_analysis} proceeds with a detailed nonlinear dynamical analysis of CM modes in model conjugated hydrocarbons for which we can perform extensive simulations. Specifically, we investigate the phase-space structures that enable periodic CM modes and establish the solitary-wave mechanism that supports such dynamics.
Then, in section~\ref{sec:CM_in_real_molecule}, we build on those results to provide evidence for the same type of attosecond-soliton CM mode in full quantum-chemistry computations of the bromohexatriyne molecule. We also discuss experimental implications of our findings.
Section~\ref{sec:conclusion} concludes the paper and discusses implication of our results for future CM studies.

\section{Theoretical and computational framework}\label{sec:CM_dynamics}

We consider field-free CM in a conjugated hydrocarbon molecular cation following the sudden creation of a localized one-electron hole in the system, {\it e.g.}, as would result from site-specific ionization.
Our general approach is sketched in figure~\ref{fig:Fig_1}: we initially put the localized hole at one end of the conjugated $\pi$ system and study how the ensuing CM dynamics moves the electron/hole 
density across the entire molecule.
The delocalized $\pi$ structure of conjugated hydrocarbons has previously been shown to serve as the backbone for CM in these compounds~\cite{Folorunso2021}.
To demonstrate the generality of our finding, we consider two levels of theory: (1) a reduced model of the $\pi$ system for which we can perform extensive computations and detailed analyses, and (2) high-performance quantum-chemistry simulations in a real molecule, where we leverage what we learned in the reduced case.
The reduced $\pi$ system corresponds to a one-dimensional (1D) model of an alkene. It emulates the  carbon chain conjugated structure by matching the bond lengths between the various C centers of the full 3D molecules.
We detail our model $\pi$ system in Appendix~\ref{app:model_pi_system}.

In all simulations we use time-dependent density-functional theory (TDDFT) with fixed nuclei.
TDDFT has been shown to successfully explain CM experiments and to reproduce correlated wave-function CM simulations~\cite{Kraus2015,Bruner2017}. More broadly, TDDFT computations are commonly used in atto/femtosecond science for their ability to systematically handle large molecules with many active and correlated electrons~\cite{Sandor2019,Tuthill2020}.
For a molecular cation with $N-1$ active electrons the TDDFT dynamics, in the Kohn-Sham (KS) formalism~\cite{Kohn1965}, is given by the system of one-particle equations
\begin{equation} \label{eq:TDDFT}
    i \partial_t \phi_k (\vec{r};t) = 
        \hat{\mathcal{H}}_\text{eff}\left[\rho\right](\vec{r}) \phi_k (\vec{r};t),
\end{equation}
together with the one-body density
\begin{equation} \label{eq:one-body_density}
    \rho(\vec{r};t) = \sum_{k=1}^{N-1}{ |\phi_k(\vec{r};t)|^2}.
\end{equation}
Following Pauli's exclusion principle, the KS orbitals $\left\{\phi_k\right\}_k$ are orthonormal wave function in their spin and space coordinates.
The one-body density provides a real-space representation of the electronic-charge distribution in the molecule. From it, the hole density $\rho_\text{h}(\vec{r};t)=\rho_N(\vec{r})-\rho(\vec{r};t)$ is computed by taking the difference with the neutral's ground-state density $\rho_N(\vec{r})$.
We stress the distinction between three related but different elements of TDDFT:
(i) The KS orbitals of equation~(\ref{eq:TDDFT}) are the \emph{dynamical variables} of the TDDFT problem; in themselves, KS orbitals do not correspond to physically observable quantities.
(ii) Molecular orbitals (MO) are \emph{time-independent} wave functions; we use them as a spatial basis set to build initial conditions in the model $\pi$ system.
(iii) The one-body density of equation~(\ref{eq:one-body_density}) is the \emph{physical observable} we are ultimately interested in for CM analyses.
We provide additional details about our TDDFT simulations in Appendix~\ref{app:TDDFT_simulations}.

In what follows, we investigate the influence of the initial hole localization on its subsequent CM dynamics. Specifically, we systematically and continuously vary the initial degree of localization of the one-electron hole around one end of the $\pi$-system structure in both the 1D and full 3D simulations.
By convention we set 0\% localization to correspond to the cation ground state and 100\% to correspond to a fully localized hole, as sketched at the top of figure~\ref{fig:Fig_1}.
Because of the different implementation of the 1D and 3D TDDFT simulations of Eqs.~(\ref{eq:TDDFT},\ref{eq:one-body_density}) (see Appendix~\ref{app:TDDFT_simulations}), we use different avenues to introduce a variably localized initial the hole in model $\pi$ system and full TDDFT computations.
We provide details about our initial hole configurations in Appendix~\ref{app:initial_hole_configuration}. 
In short, for the model $\pi$ system, we build the variably-localized hole configuration with a linear combination of a few occupied and unoccupied MOs of the corresponding cation.
To build the initial hole for the full TDDFT simulations, we use constrained DFT (cDFT)~\cite{Eshuis2009}, which combines standard energy minimization techniques with the constraint of having a certain amount of the hole localized around a specific center in the molecule. This approach allows us to consistently impose the initial-hole configuration without involving {\it ad hoc} MO mixing. The cDFT approach was shown to be successful in a previous study of CM in halocarbons~\cite{Folorunso2021}.
Following the initialization of the hole, we compute the subsequent TDDFT dynamics in the full, unrestricted TDDFT framework of equations~(\ref{eq:TDDFT},\ref{eq:one-body_density}), for both the model system and the real molecule.
Taken together, these equations correspond to a nonlinearly coupled system, which we study using tools from nonlinear dynamics.

\section{Nonlinear dynamical analysis} \label{sec:nonlinear_dynamics_analysis}

\begin{figure}[b]
    \centering
    \includegraphics[width=\linewidth]{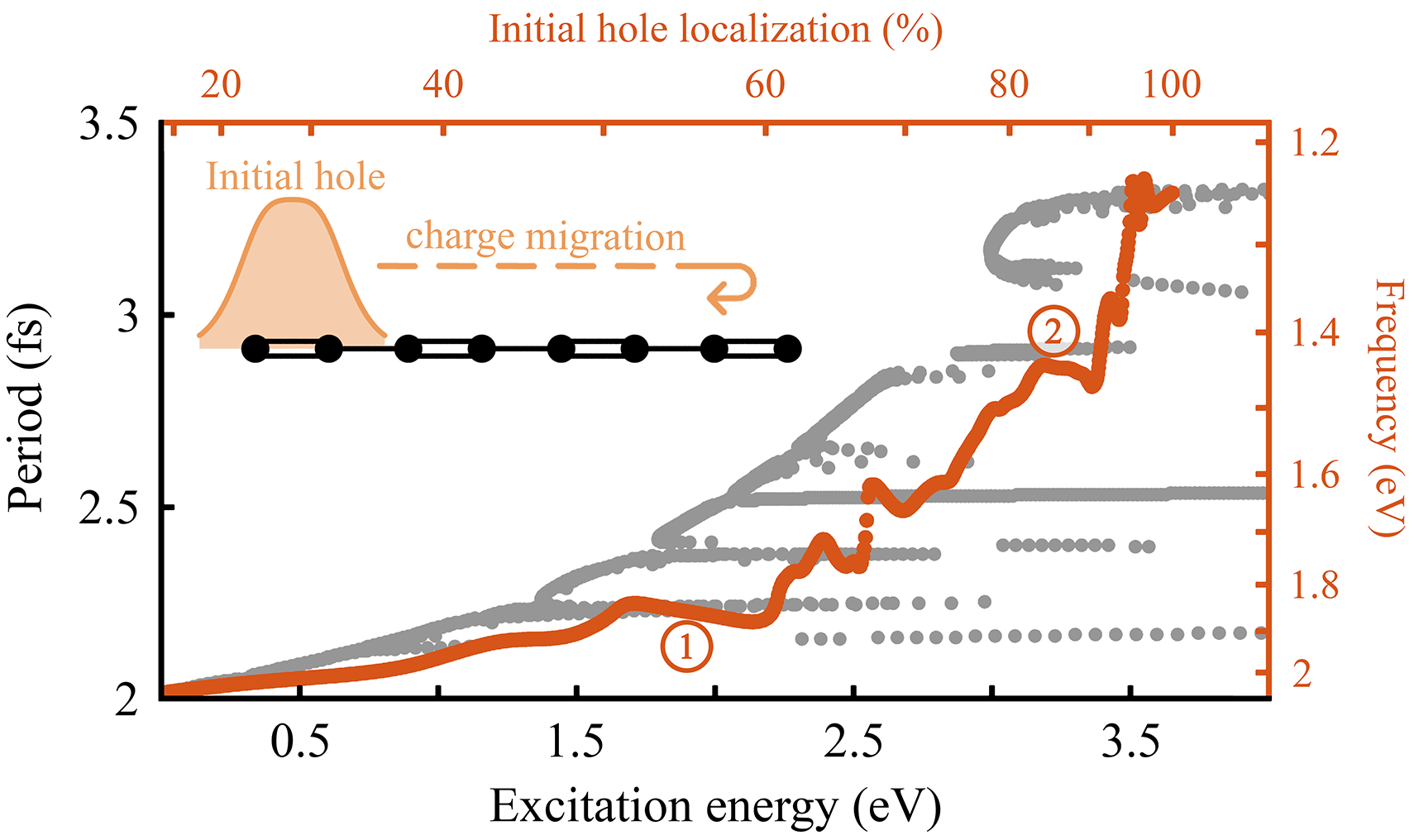}
    \caption{\label{fig:Fig_2}
    A localized one-electron hole initially introduced in the $\pi$ system of a conjugated hydrocarbon leads to CM modes whose periods vary with the initial degree of localization. 
    While varying the degree of localization of the initial hole, or equivalently the excitation energy of the molecule, we track the main frequency/period of the CM motion that moves the particle hole throughout the entire chain (dark orange dots). We compare these with direct computations of periodic CM modes in the same system (light grey dots).
    }
\end{figure}

Figure~\ref{fig:Fig_2} summarizes the result of our nonlinear analysis of CM dynamics in a reduced model of the conjugated $\pi$ system of alkene/alkyne hydrocarbons.
Specifically, we investigate the influence of the initial hole localization on the CM dynamics it induces in the molecule (dark orange markers). The more the initial hole is localized in the molecule the further away its electronic structure is from the ground-state distribution. This translates into a higher level of molecular excitation, which we indicate along the lower x axis.
For each initial hole configuration we compute the subsequent field-free CM dynamics as given by equations~(\ref{eq:TDDFT},\ref{eq:one-body_density}). We then extract the main frequency component of any motion that moves the hole density between the two ends of the molecule -- shown as the right y axis in the figure.
We aim to obtain a global picture of phase space through the dependence of this frequency on the initial conditions.
In nonlinear dynamics, this approach corresponds to a frequency-map analysis (FMA)~\cite{Laskar1993,Laskar1999} which, for instance, has been very successful in celestial mechanics to understand the dynamics inside the solar system~\cite{Laskar1993_2,Laskar1993_3}.
In particular, FMA can be used to discriminate between chaotic motions, with a strong sensitivity to the initial conditions, and regular ones, among them periodic motions.

\subsection{Frequency-map analysis}

For this work, we configure our FMA to focus on CM dynamics unfolding within the first few tens of femtoseconds following the initial localized-hole creation. 
We provide further details regarding our implementation of the FMA in Appendix~\ref{app:FMA}.
The results of the FMA, as shown in figure~\ref{fig:Fig_2}, reveal several striking features. First it shows that, as a trend, the period of the CM increases (frequency decreases) with increasing molecular excitation energy. In other words, initially more localized holes lead to slower CM. Second, this increase is irregular and proceeds in an almost step-wise fashion via a series of plateaus. Within each of these plateaus, the period of the CM is essentially independent of the excitation energy or, equivalently, of the details of the way in which the hole was initialized.

\begin{figure}[b]
    \centering
    \includegraphics[width=\linewidth]{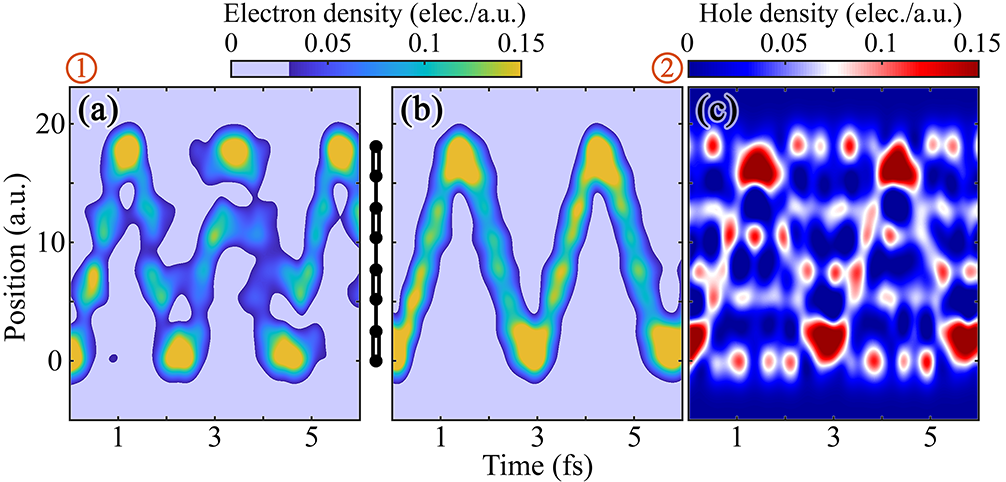}
    \caption{\label{fig:Fig_3}
    Examples of charge-migration modes in a conjugated hydrocarbon model with a particle electron/hole that periodically moves through the $\pi$ system. (a,b) Electron density contribution from the KS-orbital channel in which we introduce the initial localized-hole perturbation -- see text. (c) Hole density associated with panel (b). The overhead labels \raisebox{.5pt}{\textcircled{\raisebox{-.9pt} {1}}} and \raisebox{.5pt}{\textcircled{\raisebox{-.9pt} {2}}} indicate the plateaus of figure~\ref{fig:Fig_2} from which these are taken.}
\end{figure}

Next, we look at the type of dynamics associated with the plateaus in the FMA of figure~\ref{fig:Fig_2}.
We have found that these dynamics are best understood by looking at the density contribution from the KS orbital in which we introduce the initial localized-hole perturbation. In section~\ref{sec:KSO_vs_hole_dynamics} below we explain how this KS-orbital density matches the CM motion of the hole we are ultimately interested in.
For illustration, in figure~\ref{fig:Fig_3}~(a,b) we show the temporal evolution of two sample densities -- see the matching labels 1 and 2 in figure~\ref{fig:Fig_2}. 
Both panels reveal qualitatively similar motions where the initial electron density in the KS orbital, instead of spreading, remains localized in space and periodically propagates through the entire $\pi$ system like a particle.

\begin{figure}[h]
    \centering
    \includegraphics[width=\linewidth]{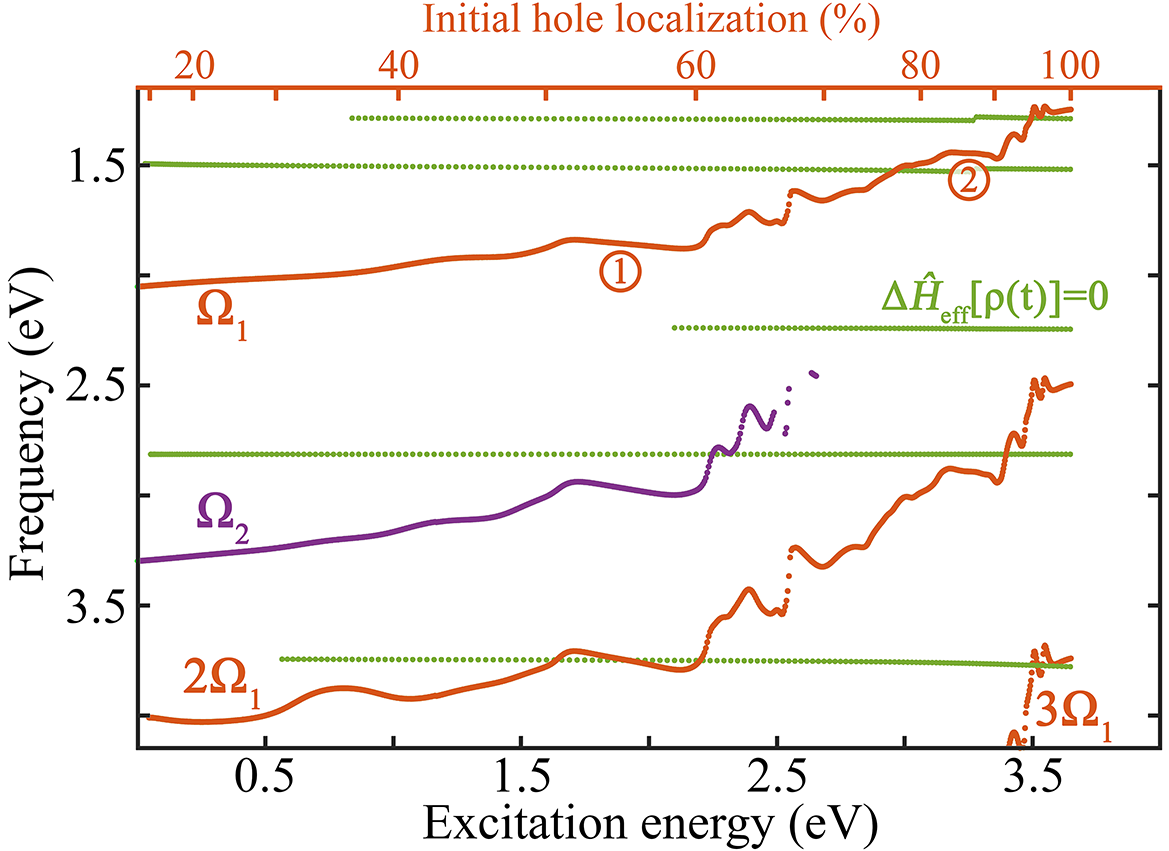}
    \caption{\label{fig:Fig_4}
    Extended frequency-map analysis (FMA) for the model $\pi$ system showing the leading frequencies of the charge-migration dynamics.
    The component labeled $\Omega_1$ corresponds to the main frequency shown in figure~\ref{fig:Fig_2}. The curves labeled $2\Omega_1$ and $3\Omega_1$ mark FMA components with twice and three times the frequency value of the main component $\Omega_1$, respectively -- note that below $\approx$3.5~eV, the $3\Omega_1$ curve plots outside of the range of display.
    The labels \raisebox{.5pt}{\textcircled{\raisebox{-.9pt} {1}}} and \raisebox{.5pt}{\textcircled{\raisebox{-.9pt} {2}}} indicate the plateaus from which the sample CM-mode of figure~\ref{fig:Fig_3} are taken.
    We also show the result of the FMA with the linearized-dynamics approximation of $\Delta\hat{\mathcal{H}}_\text{eff}=0$ in equation~(\ref{eq:TDDFT_Hamiltonian_decomposition}), which reduces the dynamics to the beating between molecular orbitals.
    }
\end{figure}

Aside from the difference in the periods, a second quantitative difference between panels (a) and (b) of figure~\ref{fig:Fig_3} is that~(b) exhibits a ``cleaner'' propagation of the electron density along the molecule. This difference in the quality of the CM motion can be explained with the extended FMA of figure~\ref{fig:Fig_4}, by looking at the next-to-leading frequency components in the CM signal.
In figure~\ref{fig:Fig_4}, in addition to the main frequency component $\Omega_1$ we identify several overtones corresponding to an anharmonic component $\Omega_2$ and two harmonic ones $2\Omega_1$ and $3\Omega_1$ (seen above $\approx$3.5~eV). 
Notably, the plateau labeled~1 has both $\Omega_1$ and $\Omega_2$ components,
which manifests as a faint aperiodicity in the evolution of the electron density. On the other hand, in plateau~2, only multiples of the lowest frequency are visible, which translates into a very clean quasi-periodic migration.
Generally speaking, figure~\ref{fig:Fig_4} shows that we have a single fundamental CM frequency plus its harmonics, and hence clean CM motion, for initial hole localization above about 60\%  in the model $\pi$ system.
We have checked that we observe qualitatively similar results in the other plateaus of the FMA.

\subsection{Kohn-Sham orbitals vs hole dynamics} \label{sec:KSO_vs_hole_dynamics}

The single KS-orbital density contributions shown in figure~\ref{fig:Fig_3}~(a,b) are important for understanding how the CM \emph{dynamics} is organized, however, recall that they do not correspond to \emph{physically} observable quantities. 
For that, in figure~\ref{fig:Fig_3}~(c) we show the hole density associated with~(b). 
Clearly, the two panels exhibit a similar pattern as the hole density mostly stays localized in space while it propagates through the $\pi$ system. Overall, the contributions from the other KS-orbital channels to the hole density show up mostly as higher-frequency patterns localized around the individual atomic centers.
We understand the correspondence between the selected KS-orbital and the hole densities as the result of the orthonormality of the KS orbitals in equation~(\ref{eq:TDDFT}): The localized density of panel~(c) corresponds to an unpaired KS-orbital channel, which therefore contributes a single electron to the one-body density of equation~(\ref{eq:one-body_density}). The other KS orbitals, which are forced to ``stay away'' from it via the orthogonality condition, all correspond to fully filled KS channels and therefore contribute two electrons to the density. In the end, the orthogonality between the unpaired and paired KS orbitals leads to the matching hole density in panel~(c)

\subsection{Solitary-wave charge migration mechanism}

To better understand the mechanism that regulates the particle-like hole dynamics shown in figure~\ref{fig:Fig_3}, we formally decompose the TDDFT Hamiltonian operator of equation~(\ref{eq:TDDFT}) into its linear and nonlinear parts
\begin{equation} \label{eq:TDDFT_Hamiltonian_decomposition}
    \hat{\mathcal{H}}_\text{eff}\left[\rho(t)\right] = 
        \hat{\mathcal{H}}_\text{eff}\left[\rho_\text{GS}\right] + 
        \Delta \hat{\mathcal{H}}_\text{eff}\left[\rho\left(t\right)\right],
\end{equation}
where $\rho_\text{GS}$ is the ground-state one-body density of the molecular cation.
Modeling the CM as a beating of MOs amounts to neglecting the nonlinear part $\Delta \hat{\mathcal{H}}_\text{eff}\left[\rho\left(t\right)\right]$ given that, in the basis of MOs, $\hat{\mathcal{H}}_\text{eff}\left[\rho_\text{GS}\right]$ is a diagonal matrix with the MO energies on its diagonal. 
We find that the linearized MO-beating approximation yields qualitatively and quantitatively different results from the full TDDFT dynamics.
To illustrate this, in figure~\ref{fig:Fig_4} we compare the extended FMA for the full TDDFT dynamics, including \emph{time-dependent} electron-electron interactions, with its linearized approximation  $\Delta\hat{\mathcal{H}}_\text{eff}=0$ in equation~(\ref{eq:TDDFT_Hamiltonian_decomposition}), which reduces the dynamics to the beating between MOs. The figure shows a clear disagreement between the two results, both in terms of the shapes of the frequency components and in the number of them.

Intuitively, we understand the inadequacy of the MO-beating picture as follows: MOs are delocalized over the entire $\pi$ system and thus, in order to obtain a tightly localized electron density over a portion of the chain, one needs the coherent superposition of multiple MO wave functions.
Then, the mismatch in the energy spacing between these MOs would lead to a decoherence, and thus spread, of the electronic density, which we do not observe in our simulations.
This shows that the dispersion associated with the linear part of $\hat{\mathcal{H}}_\text{eff}\left[\rho(t)\right]$ in equation~(\ref{eq:TDDFT_Hamiltonian_decomposition}) is cancelled by nonlinear effects associated with the nonlinear component $\Delta \hat{\mathcal{H}}_\text{eff}\left[\rho\left(t\right)\right]$, ultimately leading to the non-dispersing solitary-wave dynamics shown in figure~\ref{fig:Fig_3}.

\subsection{Periodic charge-migration soliton modes}

To conclude our nonlinear analysis of the model $\pi$ system, we return to figure~\ref{fig:Fig_2}. While the FMA of figure~\ref{fig:Fig_2} sheds light on the CM dynamics unfolding from the initial hole we impose in the conjugated system, it only explores a narrow portion of the phase space that would otherwise be accessible to CM motions in general. Note also that one could design different ways to generate the initial hole than what we have chosen here, each potentially exploring a different portion of the phase space. 
To more fully explore the phase space in our model system, we complement our FMA with a direct systematic search for periodic CM modes that exhibit a traveling solitary wave similar to figure~\ref{fig:Fig_3}~(b,c), in the full parameter space of the model $\pi$ system.
We discuss our strategy for finding these periodic CM modes in Appendix~\ref{app:periodic_CM_mode_search}.
We show the period and excitation-energy of these periodic CM modes with light grey markers in figure~\ref{fig:Fig_2}.
As a whole, these results give a very clear picture of how the dynamics of particle-like CM is organized in the phase space of the $\pi$ system: The ``ladder'' of extended plateaus again shows that essentially the same CM period can be observed over a wide range of excitation energy, often spanning more than one eV. We also recover the general trend that slower CM modes are only available for more excited electronic configurations of the target -- note the lack of periodic CM modes in the upper-left corner of the figure.

In figure~\ref{fig:Fig_2}, the comparison between the periodic CM modes and the FMA -- light grey and dark orange markers, respectively -- is stunning: All the plateaus in the latter match a set of periodic modes in the former. In other words, these periodic CM modes form the dynamical skeleton that regulates the CM we observed when creating an {\it ad hoc} localized hole at one end of the $\pi$ system. The ``cleanliness'' of that motion -- {\it e.g.}, figure~\ref{fig:Fig_3}~(a) {\it vs}~(b) -- depends on how close the initial condition puts the initial electronic configuration to a suitable periodic CM mode it can mimic. 
Altogether this suggests a two-pronged approach to CM studies: (i) asking whether the molecule of interest supports periodic soliton CM modes. If so, then (ii) identify how to tailor the ionization process to access the CM mode(s) of interest. We revisit this idea in section~\ref{sec:CM_in_real_molecule} below.

\subsection{Functionalizing the carbon chain}

The bare conjugated-carbon chain we have considered so far is, of course, highly symmetric and therefore an impractical system for generating a localized hole at one end in a realistic experimental scenario.
Instead, one can consider functionalizing the chain by attaching a functional group at one of its ends, as sketched in figure~\ref{fig:Fig_1}. In simulations, we do so both for the model $\pi$ system and full TDDFT simulations (see section~\ref{sec:CM_in_real_molecule}). 
Specifically, in this paper we consider the example of using a halogen functional group.

\begin{figure}[b]
    \centering
    \includegraphics[width=\linewidth]{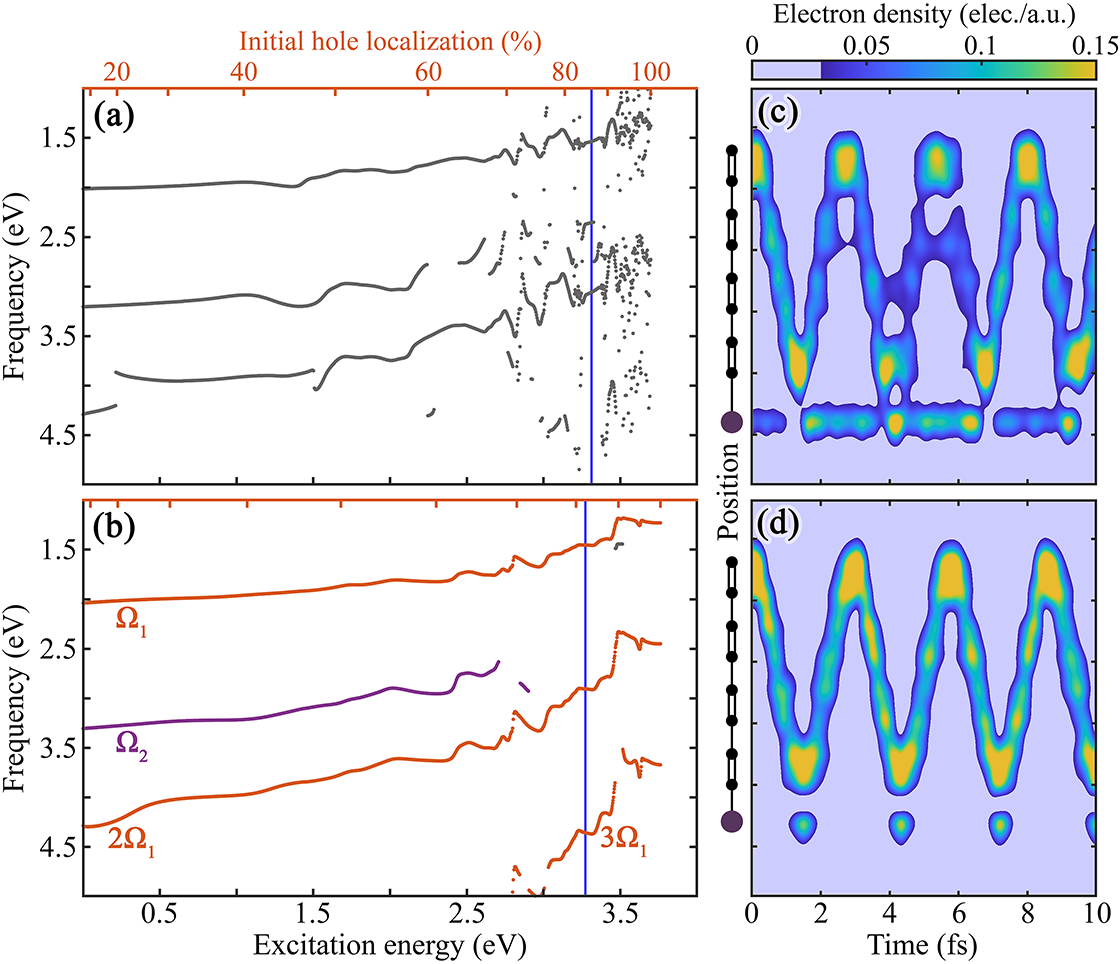}
    \caption{\label{fig:Fig_5}
    Molecular functionalization can be used to steer charge migration (CM):
    Adding different functional groups to the same carbon chain can (top) inhibit or (bottom) enhance CM modes in the rest of the $\pi$ system.
    Here we vary the functionalization by having different distances between the halogen and the chain: 4~a.u.\ in panels~(a,c) and 3~a.u.\ in~(b,d).
    The figure compares CM dynamics after initiating the variably localized hole on the chain side of the molecule. 
    Panels~(a,b) correspond to extended FMAs similar to figure~\ref{fig:Fig_4}. For the initial condition marked by the vertical blue lines, panels~(c,d) show two sample CM dynamics as in figure~\ref{fig:Fig_3}.
    In panel (b), like in figure~\ref{fig:Fig_4}, the labels $2\Omega_1$ and $3\Omega_1$ indicate FMA frequency components with twice and three times the fundamental frequency $\Omega_1$, respectively.
    }
\end{figure}

Experiments and simulations have shown that strong field ionization can create localized ionization on the halogen center~\cite{Sandor2019} in halocarbons and that the resulting valence hole can then migrate through the rest of the carbon-chain conjugated system~\cite{Kraus2015,Folorunso2021}.
In the model $\pi$ system, we can emulate a halogen function by putting a single atomic center at one end of the conjugated-carbon chain. When varying the properties of this atomic center we have found that, in order for the  electron/hole perturbation to migrate between the function and the carbon chain
as it does in our original CM example of figure~\ref{fig:Fig_1}, 
their respective orbitals must hybridize when forming the overall compound's electronic structure. 
This finding is consistent with observations in full TDDFT simulations of functionalized alkynes using different halogens~\cite{Folorunso2021}.
Intuitively, such delocalized hybridized orbitals provide a bridge for the electron density to move between the function and carbon-chain parts of the molecule. 

To conclude this section, we illustrates a second potential use of molecular functionalization, as a means to steer CM dynamics. 
In the same type of halo-functionalized model $\pi$ system as previsouly, we now start the hole on the chain side of the compound. Figure~\ref{fig:Fig_5} compares the CM dynamics for two different functionalization configurations, here controlled with the distance between the function-atom and chain. 
Panel~(a) shows a fuzzy FMA that lacks the clean solitary-wave CM-mode dynamics of the chain alone, as illustrated in panel~(c). 
On the other hand, panel~(b) exhibits clean frequency components with plateaus similar to that of the chain on its own. This is confirmed in the sample CM dynamics of panel~(c) where the particle density travels almost periodically along the conjugated backbone and only transiently ``leaks'' to the function when the migration reaches the lower end of the chain.
In both cases, the halogen function acts as a bias for the rest of the $\pi$ system by (bottom) enhancing or (top) inhibiting its CM modes.

\section{Charge migration in ``real'' conjugated molecules} \label{sec:CM_in_real_molecule}

We now leverage the results of our nonlinear analysis of CM dynamics in the model $\pi$ system and apply them to ``real'' molecules and to experimental considerations. Though we cannot systematically explore the phase space in the full molecule in the way we did in the reduced system, the FMA analysis is quite revealing in terms of what we have already observed.
We provide details of the TDDFT simulations, initial-hole configuration, and FMA computations in bromohexatriyne in Appendices~\ref{app:TDDFT_simulations}, \ref{app:initial_hole_configuration}, and~\ref{app:FMA}, respectively.

\begin{figure}[b]
    \centering
    \includegraphics[width=\linewidth]{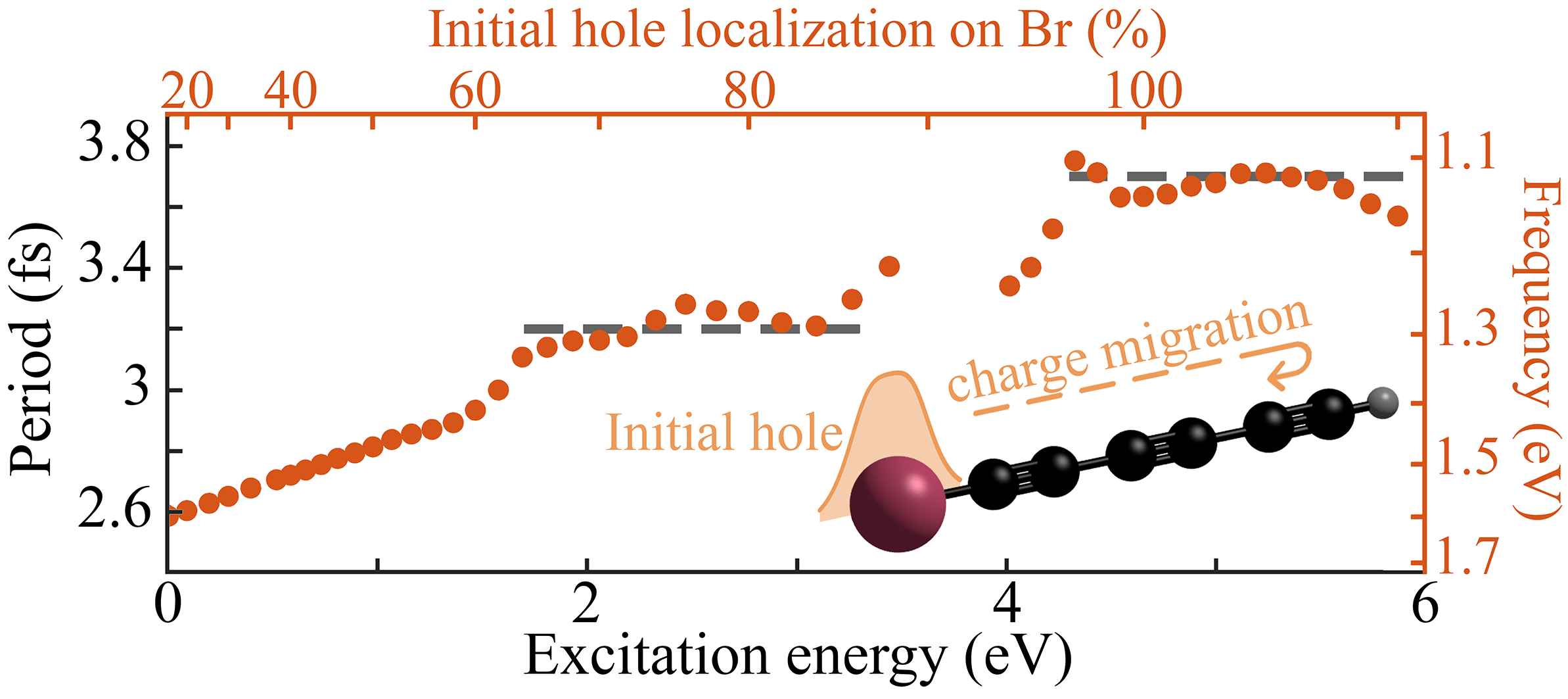}
    \caption{\label{fig:Fig_6}
    A similar charge-migration frequency/period analysis as in figure~\ref{fig:Fig_2} can be applied to full-dimension, all-electron, quantum-chemistry TDDFT simulations in bromohexatriyne (inset molecule). Orange dots mark the main frequency of the CM motion that moves the hole throughout the entire molecule -- between the Br and final C$\equiv$C-H groups. We added the horizontal dashed grey lines by hand to highlight the two plateaus we observe in the frequency map.}
\end{figure}

In figure~\ref{fig:Fig_6} we show the result of an FMA analysis of CM in the singly-ionized bromohexatriyne molecular cation computed with the high-performance full-dimensional {\it ab initio} quantum-chemistry TDDFT package NWChem~\cite{Lopata2011,Apra2020}. 
Here we vary the initial portion of the electron hole localized on the Br atom using cDFT as discussed above and, after releasing the hole-localization constraint and following the time evolution, we compute the main frequency of CM dynamics that move the density between the Br and final C$\equiv$C$-$H groups in the target -- orange markers in figure~\ref{fig:Fig_6}. The results are strikingly similar to the simplified $\pi$-system model: 
As a trend, the period of the CM increases with the initial-hole localization (excitation energy). This increase happens in a step-wise fashion over ranges of initial conditions that lead to essentially the same CM period.  Note the two distinctive plateaus between about 1.7~eV and 3.3~eV, and 4.3~eV and 6~eV, highlighted with the dashed grey lines in the figure.
We have checked that the hole dynamics in these plateaus resembles that of the model $\pi$ system -- see reference~\cite{Folorunso2021}, and its supplemental movie, for sample CM motions in full simulations of halocarbons.

A notable quantitative difference between the FMA in the model $\pi$ system of figure~\ref{fig:Fig_2} and in the full bromohexatriyne of figure~\ref{fig:Fig_6} is that the latter has apparently fewer plateaus, and that the plateaus extend over a wider range of excitation energies. We speculate on two possible mechanisms for this.
(i) \emph{ionization condition:} By using an energy-minimization condition, the cDFT might be more ``gentle'' than the {\it ad hoc} initial-hole configuration we use in the model $\pi$ system. This would lead the FMA in bromohexatriyne to effectively explore a relatively narrower portion of phase space and thus ``see'' fewer of the CM-mode plateaus that the molecule can support.
In an experimental context this speaks to the idea that, by properly tuning the initial parameters, one can have an ionization process that yields a relatively large bandwidth of excitation energy but still induces a consistent CM response.
(ii) \emph{structural condition:} Compared to the rest of the alkyne group, the Br function is effectively a heterogeneous center.
So, while the alkyne and Br components properly hybridize to form the molecular cation, this mixing might not be compatible with all the CM-mode plateaus of the bare C chain alone. In turn this would result in a net reduction of the number of plateaus supported by the molecule and thus accessible to the FMA.
This second explanation speaks to the idea of chemical control of CM motions, where a functional group added to a conjugated organic system can act as a bias and alter the ability of the rest of the molecule to support periodic CM modes
-- see the discussion around figure~\ref{fig:Fig_5}.
More broadly, the possibility for altering a molecule's ability to support CM dynamics through changes in its electronic configuration opens the door to doing so through external knobs like laser fields: The laser would selectively switch sustained CM ``on'' or ``off'' in the target sample by enabling or preventing its CM modes.

\section{Conclusion and outlook} \label{sec:conclusion}

In conclusion, we have performed detailed analyses of CM in conjugated organic molecules using tools from nonlinear dynamics (FMA and periodic-motion analysis).
In the density picture we showed that periodic CM modes, with a hole traveling back-and-forth through the $\pi$ system in a particle-like manner, emerge as solitary waves. This mechanism is fundamentally different from the few-orbital beating pictures that have previously been employed~\cite{Remacle2006,Calegari2014,Ayuso2017} and is driven by \emph{time-dependent} multi-electron interactions.
Obviously, for a detailed quantitative prediction of CM modes, like their precise period or other metrics associated with the CM dynamics~\cite{Folorunso2021}, one needs a detailed modeling of the molecular system. 
However, we showed that a lower level of theory, using a simplified conjugated model, qualitatively reproduces key features of the full system. In an extension of the results shown here we have found similar features when further simplifying our model $\pi$ system and neglecting exchange and correlation interactions in the simulations (see Appendix~\ref{app:no_XC}). This suggests that the particle-like CM motions we identify emerge as a result of the dynamical mean-field interaction alone. 
Long-range many-electron interactions are a hallmark of conjugated organic molecules, which makes the possibility for sustained CM motions widely available in these systems -- at least until the onset of nuclear motion.

Surprisingly, our analysis reveals that the same molecule can support several soliton CM modes with very different periods. For the  model $\pi$ system shown in figure~\ref{fig:Fig_2}, the identified CM  periods vary over a range of about 1~fs. In the full bromohexatriyne simulations of figure~\ref{fig:Fig_6}, the two identified CM modes are 500~as apart. 
Those modes are characterized by regions of the parameter space over which different initial conditions lead to the same almost-periodic CM motion, all essentially with the same period. In other words, the overall CM dynamics is not so much determined by the details of the individual electronic degrees of freedom -- nor the competition between them -- than it is driven by their collective response. This type of collective behavior is usually referred to as synchronization and has been identified throughout physics, engineering and biology~\cite{Acebron2005,Feldhaus2005,Pikovsky2007}.
We also note that these findings are compatible with the fully correlated Schr\"odinger-equation formalism, where field-free dynamics can only take discrete frequency values associated with the energy difference between the all-electron excited-state wave functions, and that more excited levels tend to be more closely spaced leading to longer periods.
In comparison, our attosecond-soliton mechanism provides a novel way to understand CM dynamics directly in the time domain and real space.
In all cases, the type of parametric robustness of the CM dynamics we have identified is essential for experimental applications as it provides a robustness of the migration dynamics against uncertainties in the way the hole might be created.
The results presented in this work, along with the analysis tools that we employed, can help provide important perspectives for the design of future CM studies.


\begin{acknowledgments}
We thank L.F.~DiMauro and R.R.~Jones for helpful discussions on this work.
This work was supported by the U.S.\ Department of Energy, Office of Science, Office of Basic Energy Sciences, under Award No.~DE-SC0012462.
\end{acknowledgments}

\appendix* \section*{Appendices}

\subsection{Reduced $\pi$-system model} \label{app:model_pi_system}

We systematically build our reduced $\pi$ system from one-dimensional (1D) carbon chains, denoted $\left(\text{C}_2\right)_n$, with $n$ the number of pairs of ``C'' centers. These chains are the 1D analog to 3D alkenes, without the hydrogen centers. 
We use soft-Coulomb~\cite{Javanainen1988} effective potentials to describe electron interactions, parameterized as
$
    V_{\text{sc}}\left[Z,a\right]\left(x\right) = -\frac{Z}{\sqrt{x^{2}+a^{2}}},
$
with $Z$ the effective charge and $a$ the softening parameter.
For electron-electron interactions we take a softening parameter $a_\text{ee}=\sqrt{2}$.
In the model $\pi$ system, each C center contributes one electron with $Z_\text{C}=1$. We set the respective positions for the atomic centers to emulate conjugation-like bonding by using slightly different inter-atomic distances within and between C$_{2}$ pairs, respectively set to $2.5$~a.u.\ and $2.7$~a.u. We chose these distances to be comparable to the ones from full-dimensional (functionalized) alkenes.
Finally, for electron-ion interactions we select the softening parameter $a_\text{C}=1$ for which we consistently find reasonable electronic properties throughout the $\left(\text{C}_{2}\right)_{n}$ family, both in terms of molecular orbital energies and shapes.

For simplicity, we only show results for the $\left(\text{C}_{2}\right)_{4}$ system. We systematically find similar results -- multiple plateaus in the FMA associated with periodic CM modes with a traveling particle-like hole, longer CM periods accessible only to more excited configurations, etc.\ -- for the other members of the $\left(\text{C}_{2}\right)_{n}$ family.

To functionalize the model $\pi$ system we add a single atomic ``X'' center with $Z_\text{X}=2$ at one end of the molecule, meant to emulate a halogen group. We then use the softening parameter $a_\text{X}$ and distance between the function and chain to tune the coupling/hybridization between the two.

\subsection{(TD)DFT simulations} \label{app:TDDFT_simulations}

For the 1D model $\pi$ system, we use  spin-restricted TDDFT with local-density approximation (LDA) Slater-exchange and correlation potentials~\cite{Helbig2011} and an average-density self-interaction correction (ADSIC)~\cite{Legrand2002}.
Numerically, we discretize the model $\pi$ system and simulate its TDDFT dynamics of Eq.~(\ref{eq:TDDFT}) on a Cartesian grid.

For full-dimensional ab-initio TDDFT simulations in bromohexatriyne~\cite{Folorunso2021}, we use an all-electron, spin polarized level of theory with the hybrid PBE0 functional,  cc-pVDZ basis set for the H and C atoms and Stuttgart RLC ECP for Br, as implemented in the NWChem package~\cite{Lopata2011,Apra2020}.

\subsection{Variable initial hole configuration} \label{app:initial_hole_configuration}

We initialize all simulations with a one-electron hole in the conjugated system of the molecular cation.

For the model $\pi$ system, we take an intuitive approach to building the initial condition using a single parameter to control the initial hole localization.
In this system, the molecular cation has one unpaired KS orbital, while the other ones are fully filled.
In the unpaired KS-orbital channel, we linearly mix a one-electron wave function localized around an end C dimer with the delocalized cation's ground state highest-occupied molecular orbital; the initial localization of the hole is controlled by this mixing coefficient.
We then reorthonormalize the remaining paired KS orbitals. In the end, the difference in the number of electrons contributed by the unpaired and paired KS orbital yields the relative deficit of electronic density over one of the final C dimers.

\begin{figure}
    \centering
    \includegraphics[width=\linewidth]{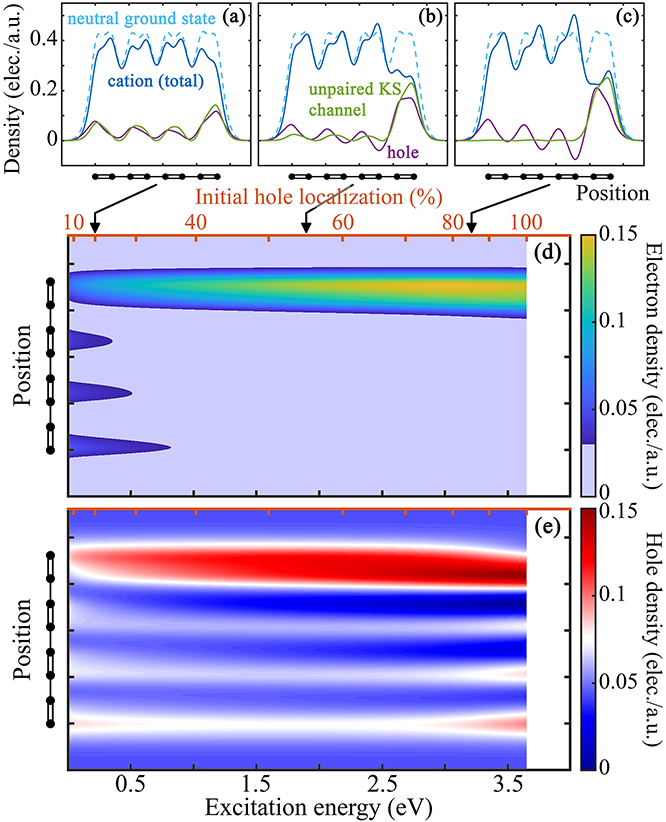}
    \caption{\label{fig:Fig_7}
    Variation of the initial electron/hole configuration we use for the frequency-map analysis (FMA) of the model $\pi$ system.
    (a-c) Samples of the initial electron and hole density distributions along the molecular backbone for 20\%, 55\%, and 85\% localization, respectively.
    (d) Initial electron density in the unpaired Kohn-Sham~(KS) orbital we use as initial condition in the FMA of figure~\ref{fig:Fig_2}.
    (e) Initial hole density associated with~(d).
    Note that panels~(d,e) use the same colormaps as in figure~\ref{fig:Fig_3}.}
\end{figure}

Figure~\ref{fig:Fig_7} illustrates how the initial electron and hole configuration depends on the localization parameter in the model $\pi$ system.
The colormaps stop at 100\% initial hole localization because our initial-condition mixing scheme has an upper bound for the excitation energy it can induce in the cation. On the other hand, we do not have the same upper limit when directly searching for periodic CM modes in the full parameter space, and we can therefore achieve the higher excitation energies shown in figure~\ref{fig:Fig_2}

\begin{figure}
    \centering
    \includegraphics[width=\linewidth]{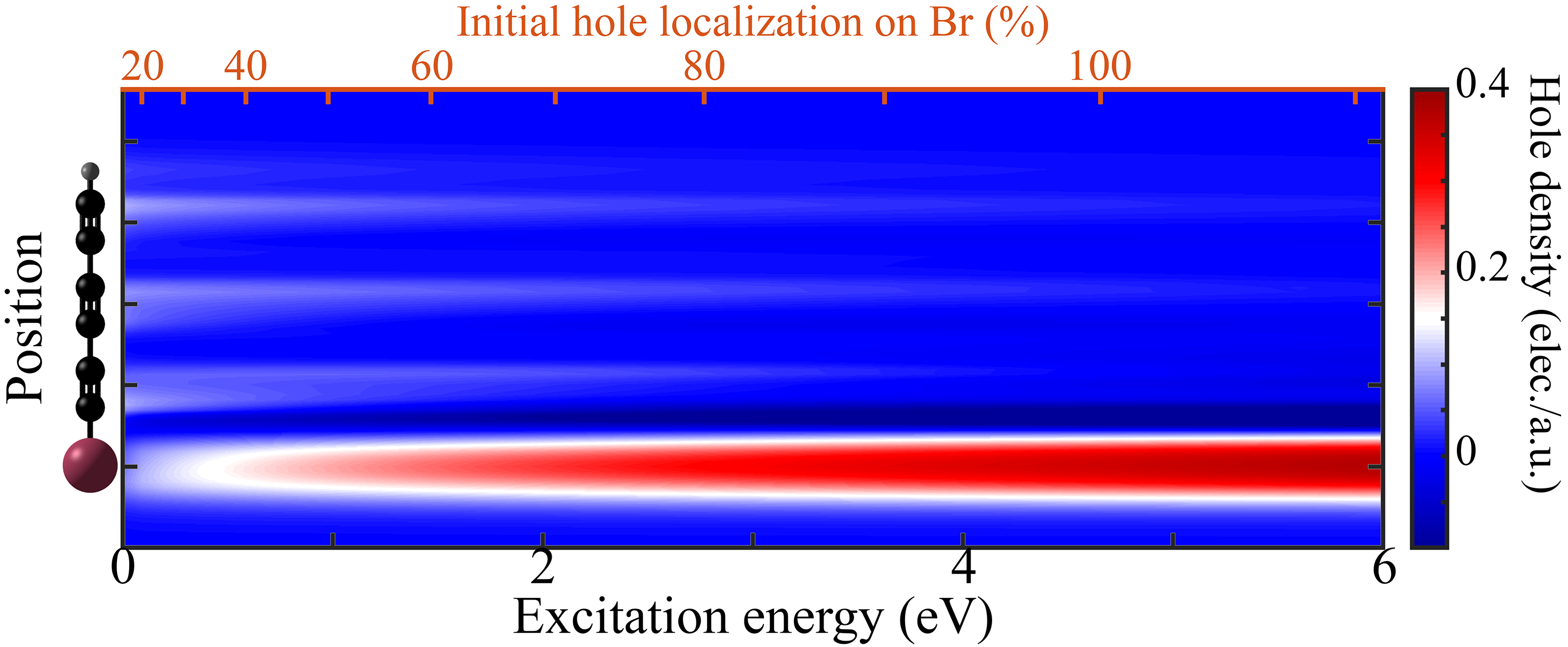}
    \caption{\label{fig:Fig_8}
    Like in figure~\ref{fig:Fig_7}~(e), variation of the initial hole configuration for the frequency-map analysis (FMA) of bromohexatriyne of figure~\ref{fig:Fig_6}. }
\end{figure}

For full (3D) TDDFT simulations, the use of the constrained DFT (cDFT) method~\cite{Eshuis2009} was described in reference~\cite{Folorunso2021}. 
In reference~~\cite{Folorunso2021} all CM simulations were initialized with the constraint of having exactly one electron hole on the halogen center. Here, instead, we keep the overall one-electron hole but vary the amount of that hole that is constrained to be localized around the Br center.
Figure~\ref{fig:Fig_8} illustrates how the initial hole configuration varies with the DFT constraint in bromohexatriyne computations.
Note that the cDFT algorithm enables us to impose over 100\% initial localization of the hole on the Br end.

\subsection{Frequency-map analysis (FMA)} \label{app:FMA}

Generally speaking, an FMA tracks the \emph{variation} of the main frequency component(s) associated with a dynamical process as a function of a continuously-varied initial condition~\cite{Laskar1993,Laskar1999}. While we use slightly different implementation of the FMA for the analysis of the model $\pi$ system and full (3D) TDDFT simulations, the underlying idea is the same in both cases: To analyze CM motions across the entire molecule, we construct a complex-valued scalar signal $s_\text{FMA}(t)= s_{l}(t)+i s_{r}(t)$ by computing the amount of electron/hole density around the left and right ends of the molecule, respectively $s_{l,r}$.
For the FMA itself, we use the time interval 10~fs $\leq t\leq$ 30~fs: We discard the first 10~fs to avoid transient effects associated with the sudden introduction of the localized perturbation in the system. We have checked that including them in the analysis has only cosmetic effects and does not change our results and conclusions. 

For the model $\pi$ system (figure~\ref{fig:Fig_2}), we compute $s_{l,r}$ by projecting the electron density contribution for the unpaired KS orbital over the end C$_2$ dimers. We choose this projection for consistency with the results of figure~\ref{fig:Fig_3}~(a,b) where we showed that the underlying organization of the CM dynamics is best apparent in this KS-orbital channel. We have checked that we obtain essentially the same results when projecting the entire hole density on the same end dimers.

For the full TDDFT simulations (figure~\ref{fig:Fig_6}), we compute $s_{l,r}$ by simply integrating the hole density around the Br (left side of the median plane to the Br-C bond) and around the -C$\equiv$C-H (right side of the median plane to the final $\equiv$C-C$\equiv$ bond) groups, respectively.

\subsection{Periodic CM mode search} \label{app:periodic_CM_mode_search}

The TDDFT system of equations~(\ref{eq:TDDFT},\ref{eq:one-body_density}) has an infinite number of degrees of freedom, which makes it impractical for a direct computation of periodic CM modes. Instead, we employ a two-step approach: 
First, we restrict the dynamics to few occupied and unoccupied MOs of the corresponding cation and use a nonlinear solver -- here the Levenberg-Marquardt Method as implemented in MATLAB$^\copyright$ -- to find periodic motions in that restricted space. 
Then, we use those restricted periodic modes as initial conditions in unrestricted TDDFT simulations and check that they still correspond to periodic motions.
We stress that, with this approach, the restricted computations only serve as an intermediary to determining periodic CM modes in the same TDDFT framework we use for our other CM simulations.

\subsection{$\pi$-system model without exchange-correlation interactions} \label{app:no_XC}

\begin{figure}
    \centering
    \includegraphics[width=\linewidth]{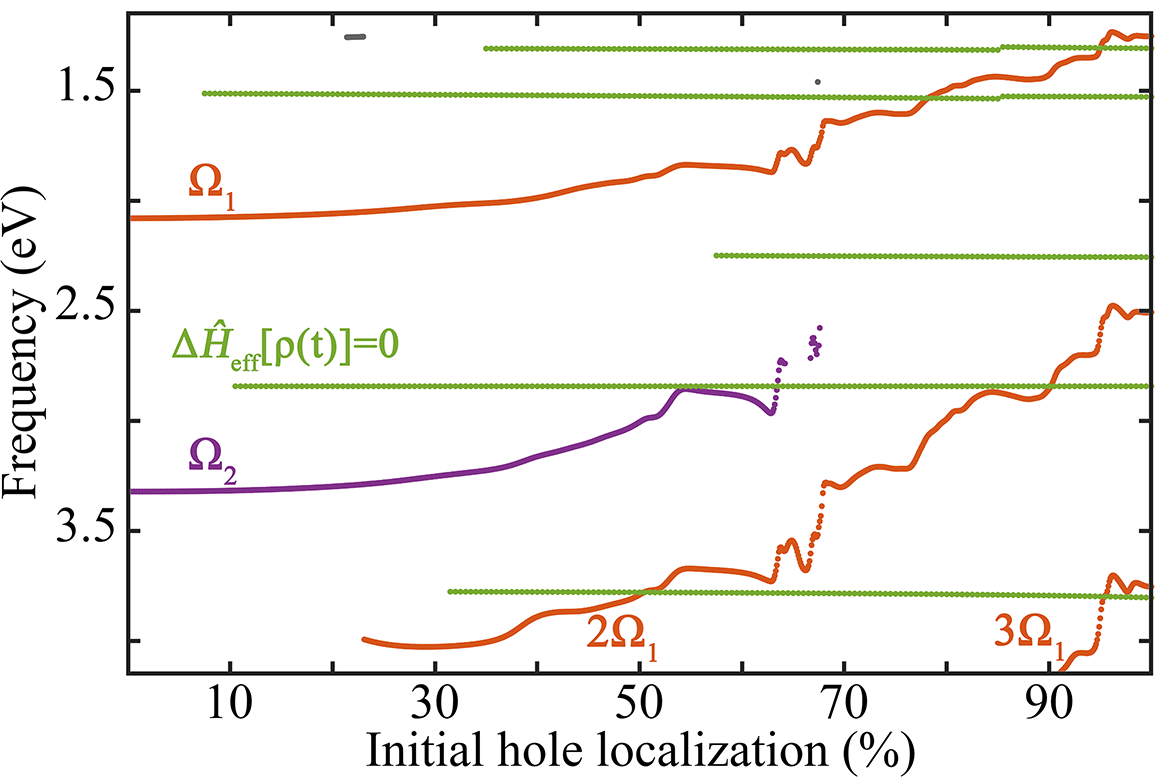}
    \caption{\label{fig:Fig_9}
    Same extended frequency-map analysis (FMA) as in figure~\ref{fig:Fig_4}, but here we have neglected the exchange-correlation term in our TDDFT simulation -- {\it i.e.}, electron-electron interactions are reduced to the mean-field Hartree potential only.
    The grey curves are the FMA components for the TDDFT simulations, labeled by the frequency values $\Omega_1$, its multiples $2\Omega_1$ and $3\Omega_1$, and $\Omega_2$. 
    The green horizontal lines show the result of an FMA starting from the same initial conditions but using the linearized-dynamics approximation of $\Delta\hat{\mathcal{H}}_\text{eff}=0$ in equation~(\ref{eq:TDDFT_Hamiltonian_decomposition}), which reduces the dynamics to the beating between molecular orbitals.}
\end{figure}

Figure~\ref{fig:Fig_9} illustrates how one obtains qualitatively very similar results when neglecting exchange-correlation interactions in the model $\pi$ system -- compare to figure~\ref{fig:Fig_4}.

\begin{thebibliography}{0}%
\makeatletter
\providecommand \@ifxundefined [1]{%
 \@ifx{#1\undefined}
}%
\providecommand \@ifnum [1]{%
 \ifnum #1\expandafter \@firstoftwo
 \else \expandafter \@secondoftwo
 \fi
}%
\providecommand \@ifx [1]{%
 \ifx #1\expandafter \@firstoftwo
 \else \expandafter \@secondoftwo
 \fi
}%
\providecommand \natexlab [1]{#1}%
\providecommand \enquote  [1]{``#1''}%
\providecommand \bibnamefont  [1]{#1}%
\providecommand \bibfnamefont [1]{#1}%
\providecommand \citenamefont [1]{#1}%
\providecommand \href@noop [0]{\@secondoftwo}%
\providecommand \href [0]{\begingroup \@sanitize@url \@href}%
\providecommand \@href[1]{\@@startlink{#1}\@@href}%
\providecommand \@@href[1]{\endgroup#1\@@endlink}%
\providecommand \@sanitize@url [0]{\catcode `\\12\catcode `\$12\catcode
  `\&12\catcode `\#12\catcode `\^12\catcode `\_12\catcode `\%12\relax}%
\providecommand \@@startlink[1]{}%
\providecommand \@@endlink[0]{}%
\providecommand \url  [0]{\begingroup\@sanitize@url \@url }%
\providecommand \@url [1]{\endgroup\@href {#1}{\urlprefix }}%
\providecommand \urlprefix  [0]{URL }%
\providecommand \Eprint [0]{\href }%
\providecommand \doibase [0]{https://doi.org/}%
\providecommand \selectlanguage [0]{\@gobble}%
\providecommand \bibinfo  [0]{\@secondoftwo}%
\providecommand \bibfield  [0]{\@secondoftwo}%
\providecommand \translation [1]{[#1]}%
\providecommand \BibitemOpen [0]{}%
\providecommand \bibitemStop [0]{}%
\providecommand \bibitemNoStop [0]{.\EOS\space}%
\providecommand \EOS [0]{\spacefactor3000\relax}%
\providecommand \BibitemShut  [1]{\csname bibitem#1\endcsname}%
\let\auto@bib@innerbib\@empty
\end{thebibliography}%


\begin{thebibliography}{47}%
\makeatletter
\providecommand \@ifxundefined [1]{%
 \@ifx{#1\undefined}
}%
\providecommand \@ifnum [1]{%
 \ifnum #1\expandafter \@firstoftwo
 \else \expandafter \@secondoftwo
 \fi
}%
\providecommand \@ifx [1]{%
 \ifx #1\expandafter \@firstoftwo
 \else \expandafter \@secondoftwo
 \fi
}%
\providecommand \natexlab [1]{#1}%
\providecommand \enquote  [1]{``#1''}%
\providecommand \bibnamefont  [1]{#1}%
\providecommand \bibfnamefont [1]{#1}%
\providecommand \citenamefont [1]{#1}%
\providecommand \href@noop [0]{\@secondoftwo}%
\providecommand \href [0]{\begingroup \@sanitize@url \@href}%
\providecommand \@href[1]{\@@startlink{#1}\@@href}%
\providecommand \@@href[1]{\endgroup#1\@@endlink}%
\providecommand \@sanitize@url [0]{\catcode `\\12\catcode `\$12\catcode
  `\&12\catcode `\#12\catcode `\^12\catcode `\_12\catcode `\%12\relax}%
\providecommand \@@startlink[1]{}%
\providecommand \@@endlink[0]{}%
\providecommand \url  [0]{\begingroup\@sanitize@url \@url }%
\providecommand \@url [1]{\endgroup\@href {#1}{\urlprefix }}%
\providecommand \urlprefix  [0]{URL }%
\providecommand \Eprint [0]{\href }%
\providecommand \doibase [0]{https://doi.org/}%
\providecommand \selectlanguage [0]{\@gobble}%
\providecommand \bibinfo  [0]{\@secondoftwo}%
\providecommand \bibfield  [0]{\@secondoftwo}%
\providecommand \translation [1]{[#1]}%
\providecommand \BibitemOpen [0]{}%
\providecommand \bibitemStop [0]{}%
\providecommand \bibitemNoStop [0]{.\EOS\space}%
\providecommand \EOS [0]{\spacefactor3000\relax}%
\providecommand \BibitemShut  [1]{\csname bibitem#1\endcsname}%
\let\auto@bib@innerbib\@empty
\bibitem [{\citenamefont {Krausz}\ and\ \citenamefont
  {Ivanov}(2009)}]{Krausz2009}%
  \BibitemOpen
  \bibfield  {author} {\bibinfo {author} {\bibfnamefont {F.}~\bibnamefont
  {Krausz}}\ and\ \bibinfo {author} {\bibfnamefont {M.}~\bibnamefont
  {Ivanov}},\ }\bibfield  {title} {\bibinfo {title} {Attosecond physics},\
  }\href@noop {} {\bibfield  {journal} {\bibinfo  {journal} {Rev. Mod. Phys.}\
  }\textbf {\bibinfo {volume} {81}},\ \bibinfo {pages} {163} (\bibinfo {year}
  {2009})}\BibitemShut {NoStop}%
\bibitem [{\citenamefont {Calegari}\ \emph {et~al.}(2016)\citenamefont
  {Calegari}, \citenamefont {Trabattoni}, \citenamefont {Palacios},
  \citenamefont {Ayuso}, \citenamefont {Castrovilli}, \citenamefont
  {Greenwood}, \citenamefont {Decleva}, \citenamefont {Mart{\'{\i}}n},\ and\
  \citenamefont {Nisoli}}]{Calegari2016}%
  \BibitemOpen
  \bibfield  {author} {\bibinfo {author} {\bibfnamefont {F.}~\bibnamefont
  {Calegari}}, \bibinfo {author} {\bibfnamefont {A.}~\bibnamefont
  {Trabattoni}}, \bibinfo {author} {\bibfnamefont {A.}~\bibnamefont
  {Palacios}}, \bibinfo {author} {\bibfnamefont {D.}~\bibnamefont {Ayuso}},
  \bibinfo {author} {\bibfnamefont {M.~C.}\ \bibnamefont {Castrovilli}},
  \bibinfo {author} {\bibfnamefont {J.~B.}\ \bibnamefont {Greenwood}}, \bibinfo
  {author} {\bibfnamefont {P.}~\bibnamefont {Decleva}}, \bibinfo {author}
  {\bibfnamefont {F.}~\bibnamefont {Mart{\'{\i}}n}},\ and\ \bibinfo {author}
  {\bibfnamefont {M.}~\bibnamefont {Nisoli}},\ }\bibfield  {title} {\bibinfo
  {title} {Charge migration induced by attosecond pulses in bio-relevant
  molecules},\ }\href@noop {} {\bibfield  {journal} {\bibinfo  {journal} {J.
  Phys. B: At. Mol. Opt. Phys.}\ }\textbf {\bibinfo {volume} {49}},\ \bibinfo
  {pages} {142001} (\bibinfo {year} {2016})}\BibitemShut {NoStop}%
\bibitem [{\citenamefont {Breidbach}\ and\ \citenamefont
  {Cederbaum}(2005)}]{Breidbach2005}%
  \BibitemOpen
  \bibfield  {author} {\bibinfo {author} {\bibfnamefont {J.}~\bibnamefont
  {Breidbach}}\ and\ \bibinfo {author} {\bibfnamefont {L.~S.}\ \bibnamefont
  {Cederbaum}},\ }\bibfield  {title} {\bibinfo {title} {Universal attosecond
  response to the removal of an electron},\ }\href@noop {} {\bibfield
  {journal} {\bibinfo  {journal} {Phys. Rev. Lett.}\ }\textbf {\bibinfo
  {volume} {94}},\ \bibinfo {pages} {033901} (\bibinfo {year}
  {2005})}\BibitemShut {NoStop}%
\bibitem [{\citenamefont {L\"unnemann}\ \emph {et~al.}(2008)\citenamefont
  {L\"unnemann}, \citenamefont {Kuleff},\ and\ \citenamefont
  {Cederbaum}}]{Lunnemann2008}%
  \BibitemOpen
  \bibfield  {author} {\bibinfo {author} {\bibfnamefont {S.}~\bibnamefont
  {L\"unnemann}}, \bibinfo {author} {\bibfnamefont {A.~I.}\ \bibnamefont
  {Kuleff}},\ and\ \bibinfo {author} {\bibfnamefont {L.~S.}\ \bibnamefont
  {Cederbaum}},\ }\bibfield  {title} {\bibinfo {title} {Ultrafast charge
  migration in 2-phenylethyl-n,n-dimethylamine},\ }\href@noop {} {\bibfield
  {journal} {\bibinfo  {journal} {Chem. Phys. Lett.}\ }\textbf {\bibinfo
  {volume} {450}},\ \bibinfo {pages} {232} (\bibinfo {year}
  {2008})}\BibitemShut {NoStop}%
\bibitem [{\citenamefont {Smirnova}\ \emph {et~al.}(2009)\citenamefont
  {Smirnova}, \citenamefont {Mairesse}, \citenamefont {Patchkovskii},
  \citenamefont {Dudovich}, \citenamefont {Villeneuve}, \citenamefont
  {Corkum},\ and\ \citenamefont {Ivanov}}]{Smirnova2009}%
  \BibitemOpen
  \bibfield  {author} {\bibinfo {author} {\bibfnamefont {O.}~\bibnamefont
  {Smirnova}}, \bibinfo {author} {\bibfnamefont {Y.}~\bibnamefont {Mairesse}},
  \bibinfo {author} {\bibfnamefont {S.}~\bibnamefont {Patchkovskii}}, \bibinfo
  {author} {\bibfnamefont {N.}~\bibnamefont {Dudovich}}, \bibinfo {author}
  {\bibfnamefont {D.}~\bibnamefont {Villeneuve}}, \bibinfo {author}
  {\bibfnamefont {P.}~\bibnamefont {Corkum}},\ and\ \bibinfo {author}
  {\bibfnamefont {M.~Y.}\ \bibnamefont {Ivanov}},\ }\bibfield  {title}
  {\bibinfo {title} {High harmonic interferometry of multi-electron dynamics in
  molecules},\ }\href@noop {} {\bibfield  {journal} {\bibinfo  {journal}
  {Nature}\ }\textbf {\bibinfo {volume} {460}},\ \bibinfo {pages} {1476}
  (\bibinfo {year} {2009})}\BibitemShut {NoStop}%
\bibitem [{\citenamefont {Calegari}\ \emph {et~al.}(2014)\citenamefont
  {Calegari}, \citenamefont {Ayuso}, \citenamefont {Trabattoni}, \citenamefont
  {Belshaw}, \citenamefont {De~Camillis}, \citenamefont {Anumula},
  \citenamefont {Frassetto}, \citenamefont {Poletto}, \citenamefont {Palacios},
  \citenamefont {Decleva}, \citenamefont {Greenwood}, \citenamefont
  {Mart{\'\i}n},\ and\ \citenamefont {Nisoli}}]{Calegari2014}%
  \BibitemOpen
  \bibfield  {author} {\bibinfo {author} {\bibfnamefont {F.}~\bibnamefont
  {Calegari}}, \bibinfo {author} {\bibfnamefont {D.}~\bibnamefont {Ayuso}},
  \bibinfo {author} {\bibfnamefont {A.}~\bibnamefont {Trabattoni}}, \bibinfo
  {author} {\bibfnamefont {L.}~\bibnamefont {Belshaw}}, \bibinfo {author}
  {\bibfnamefont {S.}~\bibnamefont {De~Camillis}}, \bibinfo {author}
  {\bibfnamefont {S.}~\bibnamefont {Anumula}}, \bibinfo {author} {\bibfnamefont
  {F.}~\bibnamefont {Frassetto}}, \bibinfo {author} {\bibfnamefont
  {L.}~\bibnamefont {Poletto}}, \bibinfo {author} {\bibfnamefont
  {A.}~\bibnamefont {Palacios}}, \bibinfo {author} {\bibfnamefont
  {P.}~\bibnamefont {Decleva}}, \bibinfo {author} {\bibfnamefont
  {J.}~\bibnamefont {Greenwood}}, \bibinfo {author} {\bibfnamefont
  {F.}~\bibnamefont {Mart{\'\i}n}},\ and\ \bibinfo {author} {\bibfnamefont
  {M.}~\bibnamefont {Nisoli}},\ }\bibfield  {title} {\bibinfo {title}
  {Ultrafast electron dynamics in phenylalanine initiated by attosecond
  pulses},\ }\href@noop {} {\bibfield  {journal} {\bibinfo  {journal}
  {Science}\ }\textbf {\bibinfo {volume} {346}},\ \bibinfo {pages} {336}
  (\bibinfo {year} {2014})}\BibitemShut {NoStop}%
\bibitem [{\citenamefont {Kraus}\ \emph {et~al.}(2015)\citenamefont {Kraus},
  \citenamefont {Mignolet}, \citenamefont {Baykusheva}, \citenamefont
  {Rupenyan}, \citenamefont {Horn{\'y}}, \citenamefont {Penka}, \citenamefont
  {Grassi}, \citenamefont {Tolstikhin}, \citenamefont {Schneider},
  \citenamefont {Jensen}, \citenamefont {Madsen}, \citenamefont {Bandrauk},
  \citenamefont {Remacle},\ and\ \citenamefont {W{\"o}rner}}]{Kraus2015}%
  \BibitemOpen
  \bibfield  {author} {\bibinfo {author} {\bibfnamefont {P.}~\bibnamefont
  {Kraus}}, \bibinfo {author} {\bibfnamefont {B.}~\bibnamefont {Mignolet}},
  \bibinfo {author} {\bibfnamefont {D.}~\bibnamefont {Baykusheva}}, \bibinfo
  {author} {\bibfnamefont {A.}~\bibnamefont {Rupenyan}}, \bibinfo {author}
  {\bibfnamefont {L.}~\bibnamefont {Horn{\'y}}}, \bibinfo {author}
  {\bibfnamefont {E.}~\bibnamefont {Penka}}, \bibinfo {author} {\bibfnamefont
  {G.}~\bibnamefont {Grassi}}, \bibinfo {author} {\bibfnamefont
  {O.}~\bibnamefont {Tolstikhin}}, \bibinfo {author} {\bibfnamefont
  {J.}~\bibnamefont {Schneider}}, \bibinfo {author} {\bibfnamefont
  {F.}~\bibnamefont {Jensen}}, \bibinfo {author} {\bibfnamefont
  {L.}~\bibnamefont {Madsen}}, \bibinfo {author} {\bibfnamefont
  {A.}~\bibnamefont {Bandrauk}}, \bibinfo {author} {\bibfnamefont
  {F.}~\bibnamefont {Remacle}},\ and\ \bibinfo {author} {\bibfnamefont
  {H.}~\bibnamefont {W{\"o}rner}},\ }\bibfield  {title} {\bibinfo {title}
  {Measurement and laser control of attosecond charge migration in ionized
  iodoacetylene},\ }\href@noop {} {\bibfield  {journal} {\bibinfo  {journal}
  {Science}\ }\textbf {\bibinfo {volume} {350}},\ \bibinfo {pages} {790}
  (\bibinfo {year} {2015})}\BibitemShut {NoStop}%
\bibitem [{\citenamefont {Kuleff}\ \emph {et~al.}(2016)\citenamefont {Kuleff},
  \citenamefont {Kryzhevoi}, \citenamefont {Pernpointner},\ and\ \citenamefont
  {Cederbaum}}]{Kuleff2016}%
  \BibitemOpen
  \bibfield  {author} {\bibinfo {author} {\bibfnamefont {A.~I.}\ \bibnamefont
  {Kuleff}}, \bibinfo {author} {\bibfnamefont {N.~V.}\ \bibnamefont
  {Kryzhevoi}}, \bibinfo {author} {\bibfnamefont {M.}~\bibnamefont
  {Pernpointner}},\ and\ \bibinfo {author} {\bibfnamefont {L.~S.}\ \bibnamefont
  {Cederbaum}},\ }\bibfield  {title} {\bibinfo {title} {Core ionization
  initiates subfemtosecond charge migration in the valence shell of
  molecules},\ }\href@noop {} {\bibfield  {journal} {\bibinfo  {journal} {Phys.
  Rev. Lett.}\ }\textbf {\bibinfo {volume} {117}},\ \bibinfo {pages} {093002}
  (\bibinfo {year} {2016})}\BibitemShut {NoStop}%
\bibitem [{\citenamefont {Remacle}\ and\ \citenamefont
  {Levine}(2006)}]{Remacle2006}%
  \BibitemOpen
  \bibfield  {author} {\bibinfo {author} {\bibfnamefont {F.}~\bibnamefont
  {Remacle}}\ and\ \bibinfo {author} {\bibfnamefont {R.~D.}\ \bibnamefont
  {Levine}},\ }\bibfield  {title} {\bibinfo {title} {An electronic time scale
  in chemistry},\ }\href@noop {} {\bibfield  {journal} {\bibinfo  {journal}
  {Proceedings of the National Academy of Sciences}\ }\textbf {\bibinfo
  {volume} {103}},\ \bibinfo {pages} {6793} (\bibinfo {year}
  {2006})}\BibitemShut {NoStop}%
\bibitem [{\citenamefont {L\'{e}pine}\ \emph {et~al.}(2014)\citenamefont
  {L\'{e}pine}, \citenamefont {Ivanov},\ and\ \citenamefont
  {Vrakking}}]{Lepine2014}%
  \BibitemOpen
  \bibfield  {author} {\bibinfo {author} {\bibfnamefont {F.}~\bibnamefont
  {L\'{e}pine}}, \bibinfo {author} {\bibfnamefont {M.~Y.}\ \bibnamefont
  {Ivanov}},\ and\ \bibinfo {author} {\bibfnamefont {M.~J.~J.}\ \bibnamefont
  {Vrakking}},\ }\bibfield  {title} {\bibinfo {title} {Attosecond molecular
  dynamics: fact or fiction?},\ }\href@noop {} {\bibfield  {journal} {\bibinfo
  {journal} {Nat. Photon.}\ }\textbf {\bibinfo {volume} {8}},\ \bibinfo {pages}
  {1749} (\bibinfo {year} {2014})}\BibitemShut {NoStop}%
\bibitem [{\citenamefont {Despr\'{e}}\ \emph {et~al.}(2015)\citenamefont
  {Despr\'{e}}, \citenamefont {Marciniak}, \citenamefont {Loriot},
  \citenamefont {Galbraith}, \citenamefont {Rouz\'{e}e}, \citenamefont
  {Vrakking}, \citenamefont {L\'{e}pine},\ and\ \citenamefont
  {Kuleff}}]{Despre2015}%
  \BibitemOpen
  \bibfield  {author} {\bibinfo {author} {\bibfnamefont {V.}~\bibnamefont
  {Despr\'{e}}}, \bibinfo {author} {\bibfnamefont {A.}~\bibnamefont
  {Marciniak}}, \bibinfo {author} {\bibfnamefont {V.}~\bibnamefont {Loriot}},
  \bibinfo {author} {\bibfnamefont {M.~C.~E.}\ \bibnamefont {Galbraith}},
  \bibinfo {author} {\bibfnamefont {A.}~\bibnamefont {Rouz\'{e}e}}, \bibinfo
  {author} {\bibfnamefont {M.~J.~J.}\ \bibnamefont {Vrakking}}, \bibinfo
  {author} {\bibfnamefont {F.}~\bibnamefont {L\'{e}pine}},\ and\ \bibinfo
  {author} {\bibfnamefont {A.~I.}\ \bibnamefont {Kuleff}},\ }\bibfield  {title}
  {\bibinfo {title} {Attosecond hole migration in benzene molecules surviving
  nuclear motion},\ }\href@noop {} {\bibfield  {journal} {\bibinfo  {journal}
  {J. Chem. Phys. Lett.}\ }\textbf {\bibinfo {volume} {6}},\ \bibinfo {pages}
  {426} (\bibinfo {year} {2015})}\BibitemShut {NoStop}%
\bibitem [{\citenamefont {Weinkauf}\ \emph {et~al.}(1997)\citenamefont
  {Weinkauf}, \citenamefont {Schlag}, \citenamefont {Martinez},\ and\
  \citenamefont {Levine}}]{Weinkauf1997}%
  \BibitemOpen
  \bibfield  {author} {\bibinfo {author} {\bibfnamefont {R.}~\bibnamefont
  {Weinkauf}}, \bibinfo {author} {\bibfnamefont {E.~W.}\ \bibnamefont
  {Schlag}}, \bibinfo {author} {\bibfnamefont {T.~J.}\ \bibnamefont
  {Martinez}},\ and\ \bibinfo {author} {\bibfnamefont {R.~D.}\ \bibnamefont
  {Levine}},\ }\bibfield  {title} {\bibinfo {title} {Nonstationary electronic
  states and site-selective reactivity},\ }\href@noop {} {\bibfield  {journal}
  {\bibinfo  {journal} {J. Phys. Chem. A}\ }\textbf {\bibinfo {volume} {101}}
  (\bibinfo {year} {1997})}\BibitemShut {NoStop}%
\bibitem [{\citenamefont {Remacle}\ \emph {et~al.}(1998)\citenamefont
  {Remacle}, \citenamefont {Levine},\ and\ \citenamefont
  {Ratner}}]{Remacle1998}%
  \BibitemOpen
  \bibfield  {author} {\bibinfo {author} {\bibfnamefont {F.}~\bibnamefont
  {Remacle}}, \bibinfo {author} {\bibfnamefont {R.}~\bibnamefont {Levine}},\
  and\ \bibinfo {author} {\bibfnamefont {M.}~\bibnamefont {Ratner}},\
  }\bibfield  {title} {\bibinfo {title} {Charge directed reactivity: a simple
  electronic model, exhibiting site selectivity, for the dissociation of
  ions},\ }\href@noop {} {\bibfield  {journal} {\bibinfo  {journal} {Chem.
  Phys. Lett.}\ }\textbf {\bibinfo {volume} {285}},\ \bibinfo {pages} {25}
  (\bibinfo {year} {1998})}\BibitemShut {NoStop}%
\bibitem [{\citenamefont {Lara-Astiaso}\ \emph {et~al.}(2018)\citenamefont
  {Lara-Astiaso}, \citenamefont {Galli}, \citenamefont {Trabattoni},
  \citenamefont {Palacios}, \citenamefont {Ayuso}, \citenamefont {Frassetto},
  \citenamefont {Poletto}, \citenamefont {De~Camillis}, \citenamefont
  {Greenwood}, \citenamefont {Decleva}, \citenamefont {Tavernelli},
  \citenamefont {Calegari}, \citenamefont {Nisoli},\ and\ \citenamefont
  {Mart\'{i}n}}]{Lara-Astiaso2018}%
  \BibitemOpen
  \bibfield  {author} {\bibinfo {author} {\bibfnamefont {M.}~\bibnamefont
  {Lara-Astiaso}}, \bibinfo {author} {\bibfnamefont {M.}~\bibnamefont {Galli}},
  \bibinfo {author} {\bibfnamefont {A.}~\bibnamefont {Trabattoni}}, \bibinfo
  {author} {\bibfnamefont {A.}~\bibnamefont {Palacios}}, \bibinfo {author}
  {\bibfnamefont {D.}~\bibnamefont {Ayuso}}, \bibinfo {author} {\bibfnamefont
  {F.}~\bibnamefont {Frassetto}}, \bibinfo {author} {\bibfnamefont
  {L.}~\bibnamefont {Poletto}}, \bibinfo {author} {\bibfnamefont
  {S.}~\bibnamefont {De~Camillis}}, \bibinfo {author} {\bibfnamefont
  {J.}~\bibnamefont {Greenwood}}, \bibinfo {author} {\bibfnamefont
  {P.}~\bibnamefont {Decleva}}, \bibinfo {author} {\bibfnamefont
  {I.}~\bibnamefont {Tavernelli}}, \bibinfo {author} {\bibfnamefont
  {F.}~\bibnamefont {Calegari}}, \bibinfo {author} {\bibfnamefont
  {M.}~\bibnamefont {Nisoli}},\ and\ \bibinfo {author} {\bibfnamefont
  {F.}~\bibnamefont {MartÃ­n}},\ }\bibfield  {title} {\bibinfo {title}
  {Attosecond pump-probe spectroscopy of charge dynamics in tryptophan},\
  }\href@noop {} {\bibfield  {journal} {\bibinfo  {journal} {J. Phys. Chem.
  Lett.}\ }\textbf {\bibinfo {volume} {9}},\ \bibinfo {pages} {4570} (\bibinfo
  {year} {2018})}\BibitemShut {NoStop}%
\bibitem [{\citenamefont {M{\aa}nsson}\ \emph {et~al.}(2021)\citenamefont
  {M{\aa}nsson}, \citenamefont {Latini}, \citenamefont {Covito}, \citenamefont
  {Wanie}, \citenamefont {Galli}, \citenamefont {Perfetto}, \citenamefont
  {Stefanucci}, \citenamefont {H\"{u}bener}, \citenamefont {De~Giovannini},
  \citenamefont {Castrovilli}, \citenamefont {Trabattoni}, \citenamefont
  {Frassetto}, \citenamefont {Poletto}, \citenamefont {Greenwood},
  \citenamefont {L\'{e}gar\'{e}}, \citenamefont {Nisoli}, \citenamefont
  {Rubio},\ and\ \citenamefont {Calegari}}]{Mansson2021}%
  \BibitemOpen
  \bibfield  {author} {\bibinfo {author} {\bibfnamefont {E.~P.}\ \bibnamefont
  {M{\aa}nsson}}, \bibinfo {author} {\bibfnamefont {S.}~\bibnamefont {Latini}},
  \bibinfo {author} {\bibfnamefont {F.}~\bibnamefont {Covito}}, \bibinfo
  {author} {\bibfnamefont {V.}~\bibnamefont {Wanie}}, \bibinfo {author}
  {\bibfnamefont {M.}~\bibnamefont {Galli}}, \bibinfo {author} {\bibfnamefont
  {E.}~\bibnamefont {Perfetto}}, \bibinfo {author} {\bibfnamefont
  {G.}~\bibnamefont {Stefanucci}}, \bibinfo {author} {\bibfnamefont
  {H.}~\bibnamefont {H\"{u}bener}}, \bibinfo {author} {\bibfnamefont
  {U.}~\bibnamefont {De~Giovannini}}, \bibinfo {author} {\bibfnamefont {M.~C.}\
  \bibnamefont {Castrovilli}}, \bibinfo {author} {\bibfnamefont
  {A.}~\bibnamefont {Trabattoni}}, \bibinfo {author} {\bibfnamefont
  {F.}~\bibnamefont {Frassetto}}, \bibinfo {author} {\bibfnamefont
  {L.}~\bibnamefont {Poletto}}, \bibinfo {author} {\bibfnamefont {J.~B.}\
  \bibnamefont {Greenwood}}, \bibinfo {author} {\bibfnamefont {F.}~\bibnamefont
  {L\'{e}gar\'{e}}}, \bibinfo {author} {\bibfnamefont {M.}~\bibnamefont
  {Nisoli}}, \bibinfo {author} {\bibfnamefont {A.}~\bibnamefont {Rubio}},\ and\
  \bibinfo {author} {\bibfnamefont {F.}~\bibnamefont {Calegari}},\ }\bibfield
  {title} {\bibinfo {title} {Real-time observation of a correlation-driven sub
  3 fs charge migration in ionised adenine},\ }\href@noop {} {\bibfield
  {journal} {\bibinfo  {journal} {Comm. Chem.}\ }\textbf {\bibinfo {volume}
  {4}},\ \bibinfo {pages} {73} (\bibinfo {year} {2021})}\BibitemShut {NoStop}%
\bibitem [{\citenamefont {Vacher}\ \emph {et~al.}(2017)\citenamefont {Vacher},
  \citenamefont {Bearpark}, \citenamefont {Robb},\ and\ \citenamefont
  {Malhado}}]{Vacher2017}%
  \BibitemOpen
  \bibfield  {author} {\bibinfo {author} {\bibfnamefont {M.}~\bibnamefont
  {Vacher}}, \bibinfo {author} {\bibfnamefont {M.~J.}\ \bibnamefont
  {Bearpark}}, \bibinfo {author} {\bibfnamefont {M.~A.}\ \bibnamefont {Robb}},\
  and\ \bibinfo {author} {\bibfnamefont {J.~P.}\ \bibnamefont {Malhado}},\
  }\bibfield  {title} {\bibinfo {title} {Electron dynamics upon ionization of
  polyatomic molecules: Coupling to quantum nuclear motion and decoherence},\
  }\href@noop {} {\bibfield  {journal} {\bibinfo  {journal} {Phys. Rev. Lett.}\
  }\textbf {\bibinfo {volume} {118}},\ \bibinfo {pages} {083001} (\bibinfo
  {year} {2017})}\BibitemShut {NoStop}%
\bibitem [{\citenamefont {Despr\'e}\ \emph {et~al.}(2018)\citenamefont
  {Despr\'e}, \citenamefont {Golubev},\ and\ \citenamefont
  {Kuleff}}]{Despre2018}%
  \BibitemOpen
  \bibfield  {author} {\bibinfo {author} {\bibfnamefont {V.}~\bibnamefont
  {Despr\'e}}, \bibinfo {author} {\bibfnamefont {N.~V.}\ \bibnamefont
  {Golubev}},\ and\ \bibinfo {author} {\bibfnamefont {A.~I.}\ \bibnamefont
  {Kuleff}},\ }\bibfield  {title} {\bibinfo {title} {Charge migration in
  propiolic acid: A full quantum dynamical study},\ }\href@noop {} {\bibfield
  {journal} {\bibinfo  {journal} {Phys. Rev. Lett.}\ }\textbf {\bibinfo
  {volume} {121}},\ \bibinfo {pages} {203002} (\bibinfo {year}
  {2018})}\BibitemShut {NoStop}%
\bibitem [{\citenamefont {Ayuso}\ \emph {et~al.}(2017)\citenamefont {Ayuso},
  \citenamefont {Palacios}, \citenamefont {Decleva},\ and\ \citenamefont
  {Mart\'{i}n}}]{Ayuso2017}%
  \BibitemOpen
  \bibfield  {author} {\bibinfo {author} {\bibfnamefont {D.}~\bibnamefont
  {Ayuso}}, \bibinfo {author} {\bibfnamefont {A.}~\bibnamefont {Palacios}},
  \bibinfo {author} {\bibfnamefont {P.}~\bibnamefont {Decleva}},\ and\ \bibinfo
  {author} {\bibfnamefont {F.}~\bibnamefont {Mart\'{i}n}},\ }\bibfield  {title}
  {\bibinfo {title} {Ultrafast charge dynamics in glycine induced by attosecond
  pulses},\ }\href@noop {} {\bibfield  {journal} {\bibinfo  {journal} {Phys.
  Chem. Chem. Phys.}\ }\textbf {\bibinfo {volume} {19}},\ \bibinfo {pages}
  {19767} (\bibinfo {year} {2017})}\BibitemShut {NoStop}%
\bibitem [{\citenamefont {Bruner}\ \emph {et~al.}(2017)\citenamefont {Bruner},
  \citenamefont {Hernandez}, \citenamefont {Mauger}, \citenamefont {Abanador},
  \citenamefont {LaMaster}, \citenamefont {Gaarde}, \citenamefont {Schafer},\
  and\ \citenamefont {Lopata}}]{Bruner2017}%
  \BibitemOpen
  \bibfield  {author} {\bibinfo {author} {\bibfnamefont {A.}~\bibnamefont
  {Bruner}}, \bibinfo {author} {\bibfnamefont {S.}~\bibnamefont {Hernandez}},
  \bibinfo {author} {\bibfnamefont {F.}~\bibnamefont {Mauger}}, \bibinfo
  {author} {\bibfnamefont {P.~M.}\ \bibnamefont {Abanador}}, \bibinfo {author}
  {\bibfnamefont {D.~J.}\ \bibnamefont {LaMaster}}, \bibinfo {author}
  {\bibfnamefont {M.~B.}\ \bibnamefont {Gaarde}}, \bibinfo {author}
  {\bibfnamefont {K.~J.}\ \bibnamefont {Schafer}},\ and\ \bibinfo {author}
  {\bibfnamefont {K.}~\bibnamefont {Lopata}},\ }\bibfield  {title} {\bibinfo
  {title} {Attosecond charge migration with {TDDFT}: Accurate dynamics from a
  well-defined initial state},\ }\href@noop {} {\bibfield  {journal} {\bibinfo
  {journal} {J. Phys. Chem. Lett.}\ }\textbf {\bibinfo {volume} {8}},\ \bibinfo
  {pages} {3991} (\bibinfo {year} {2017})}\BibitemShut {NoStop}%
\bibitem [{\citenamefont {Jia}\ \emph {et~al.}(2017)\citenamefont {Jia},
  \citenamefont {Manz}, \citenamefont {Paulus}, \citenamefont {Pohl},
  \citenamefont {Tremblay},\ and\ \citenamefont {Yang}}]{Jia2017}%
  \BibitemOpen
  \bibfield  {author} {\bibinfo {author} {\bibfnamefont {D.}~\bibnamefont
  {Jia}}, \bibinfo {author} {\bibfnamefont {J.}~\bibnamefont {Manz}}, \bibinfo
  {author} {\bibfnamefont {B.}~\bibnamefont {Paulus}}, \bibinfo {author}
  {\bibfnamefont {V.}~\bibnamefont {Pohl}}, \bibinfo {author} {\bibfnamefont
  {J.~C.}\ \bibnamefont {Tremblay}},\ and\ \bibinfo {author} {\bibfnamefont
  {Y.}~\bibnamefont {Yang}},\ }\bibfield  {title} {\bibinfo {title} {Quantum
  control of electronic fluxes during adiabatic attosecond charge migration in
  degenerate superposition states of benzene},\ }\href@noop {} {\bibfield
  {journal} {\bibinfo  {journal} {Chem. Phys.}\ }\textbf {\bibinfo {volume}
  {482}},\ \bibinfo {pages} {146} (\bibinfo {year} {2017})}\BibitemShut
  {NoStop}%
\bibitem [{\citenamefont {Perfetto}\ \emph {et~al.}(2018)\citenamefont
  {Perfetto}, \citenamefont {Sangalli}, \citenamefont {Marini},\ and\
  \citenamefont {Stefanucci}}]{Perfetto2018}%
  \BibitemOpen
  \bibfield  {author} {\bibinfo {author} {\bibfnamefont {E.}~\bibnamefont
  {Perfetto}}, \bibinfo {author} {\bibfnamefont {D.}~\bibnamefont {Sangalli}},
  \bibinfo {author} {\bibfnamefont {A.}~\bibnamefont {Marini}},\ and\ \bibinfo
  {author} {\bibfnamefont {G.}~\bibnamefont {Stefanucci}},\ }\bibfield  {title}
  {\bibinfo {title} {Ultrafast charge migration in {XUV} photoexcited
  phenylalanine: A first-principles study based on real-time nonequilibrium
  {G}reen's functions},\ }\href@noop {} {\bibfield  {journal} {\bibinfo
  {journal} {J. Phys. Chem. Lett.}\ }\textbf {\bibinfo {volume} {9}},\ \bibinfo
  {pages} {1353} (\bibinfo {year} {2018})}\BibitemShut {NoStop}%
\bibitem [{\citenamefont {Folorunso}\ \emph {et~al.}(2021)\citenamefont
  {Folorunso}, \citenamefont {Bruner}, \citenamefont {Mauger}, \citenamefont
  {Hamer}, \citenamefont {Hernandez}, \citenamefont {Jones}, \citenamefont
  {DiMauro}, \citenamefont {Gaarde}, \citenamefont {Schafer},\ and\
  \citenamefont {Lopata}}]{Folorunso2021}%
  \BibitemOpen
  \bibfield  {author} {\bibinfo {author} {\bibfnamefont {A.~S.}\ \bibnamefont
  {Folorunso}}, \bibinfo {author} {\bibfnamefont {A.}~\bibnamefont {Bruner}},
  \bibinfo {author} {\bibfnamefont {F.}~\bibnamefont {Mauger}}, \bibinfo
  {author} {\bibfnamefont {K.~A.}\ \bibnamefont {Hamer}}, \bibinfo {author}
  {\bibfnamefont {S.}~\bibnamefont {Hernandez}}, \bibinfo {author}
  {\bibfnamefont {R.~R.}\ \bibnamefont {Jones}}, \bibinfo {author}
  {\bibfnamefont {L.~F.}\ \bibnamefont {DiMauro}}, \bibinfo {author}
  {\bibfnamefont {M.~B.}\ \bibnamefont {Gaarde}}, \bibinfo {author}
  {\bibfnamefont {K.~J.}\ \bibnamefont {Schafer}},\ and\ \bibinfo {author}
  {\bibfnamefont {K.}~\bibnamefont {Lopata}},\ }\bibfield  {title} {\bibinfo
  {title} {Molecular modes of attosecond charge migration},\ }\href@noop {}
  {\bibfield  {journal} {\bibinfo  {journal} {Phys. Rev. Lett.}\ }\textbf
  {\bibinfo {volume} {126}},\ \bibinfo {pages} {133002} (\bibinfo {year}
  {2021})}\BibitemShut {NoStop}%
\bibitem [{\citenamefont {Masoliver}\ and\ \citenamefont
  {Ros}(2011)}]{Masoliver2011}%
  \BibitemOpen
  \bibfield  {author} {\bibinfo {author} {\bibfnamefont {J.}~\bibnamefont
  {Masoliver}}\ and\ \bibinfo {author} {\bibfnamefont {A.}~\bibnamefont
  {Ros}},\ }\bibfield  {title} {\bibinfo {title} {Integrability and chaos: the
  classical uncertainty},\ }\href@noop {} {\bibfield  {journal} {\bibinfo
  {journal} {Eur. J. Phys.}\ }\textbf {\bibinfo {volume} {32}},\ \bibinfo
  {pages} {431} (\bibinfo {year} {2011})}\BibitemShut {NoStop}%
\bibitem [{\citenamefont {Lichtenberg}\ and\ \citenamefont
  {Lieberman}(1992)}]{RegrAndChaosDyn}%
  \BibitemOpen
  \bibfield  {author} {\bibinfo {author} {\bibfnamefont {A.}~\bibnamefont
  {Lichtenberg}}\ and\ \bibinfo {author} {\bibfnamefont {M.}~\bibnamefont
  {Lieberman}},\ }\href@noop {} {\emph {\bibinfo {title} {Regular and Chaotic
  Dynamics}}},\ Vol.~\bibinfo {volume} {38}\ (\bibinfo  {publisher}
  {Springer-Verlag},\ \bibinfo {address} {New York},\ \bibinfo {year}
  {1992})\BibitemShut {NoStop}%
\bibitem [{\citenamefont {Miller}(1998)}]{Miller1998}%
  \BibitemOpen
  \bibfield  {author} {\bibinfo {author} {\bibfnamefont {W.~H.}\ \bibnamefont
  {Miller}},\ }\bibfield  {title} {\bibinfo {title} {Spiers memorial lecture
  quantum and semiclassical theory of chemical reaction rates},\ }\href@noop {}
  {\bibfield  {journal} {\bibinfo  {journal} {Faraday Discuss.}\ }\textbf
  {\bibinfo {volume} {110}},\ \bibinfo {pages} {1} (\bibinfo {year}
  {1998})}\BibitemShut {NoStop}%
\bibitem [{\citenamefont {Bartsch}\ \emph {et~al.}(2005)\citenamefont
  {Bartsch}, \citenamefont {Hernandez},\ and\ \citenamefont
  {Uzer}}]{Bartsch2005}%
  \BibitemOpen
  \bibfield  {author} {\bibinfo {author} {\bibfnamefont {T.}~\bibnamefont
  {Bartsch}}, \bibinfo {author} {\bibfnamefont {R.}~\bibnamefont {Hernandez}},\
  and\ \bibinfo {author} {\bibfnamefont {T.}~\bibnamefont {Uzer}},\ }\bibfield
  {title} {\bibinfo {title} {Transition state in a noisy environment},\
  }\href@noop {} {\bibfield  {journal} {\bibinfo  {journal} {Phys. Rev. Lett.}\
  }\textbf {\bibinfo {volume} {95}},\ \bibinfo {pages} {058301} (\bibinfo
  {year} {2005})}\BibitemShut {NoStop}%
\bibitem [{\citenamefont {Kawai}\ \emph {et~al.}(2007)\citenamefont {Kawai},
  \citenamefont {Bandrauk}, \citenamefont {Jaff\'{e}}, \citenamefont {Bartsch},
  \citenamefont {Palaci\'{a}n},\ and\ \citenamefont {Uzer}}]{Kawai2007}%
  \BibitemOpen
  \bibfield  {author} {\bibinfo {author} {\bibfnamefont {S.}~\bibnamefont
  {Kawai}}, \bibinfo {author} {\bibfnamefont {A.~D.}\ \bibnamefont {Bandrauk}},
  \bibinfo {author} {\bibfnamefont {C.}~\bibnamefont {Jaff\'{e}}}, \bibinfo
  {author} {\bibfnamefont {T.}~\bibnamefont {Bartsch}}, \bibinfo {author}
  {\bibfnamefont {J.}~\bibnamefont {Palaci\'{a}n}},\ and\ \bibinfo {author}
  {\bibfnamefont {T.}~\bibnamefont {Uzer}},\ }\bibfield  {title} {\bibinfo
  {title} {Transition state theory for laser-driven reactions},\ }\href@noop {}
  {\bibfield  {journal} {\bibinfo  {journal} {J. Chem. Phys.}\ }\textbf
  {\bibinfo {volume} {126}},\ \bibinfo {pages} {164306} (\bibinfo {year}
  {2007})}\BibitemShut {NoStop}%
\bibitem [{\citenamefont {van~de Sand}\ and\ \citenamefont
  {Rost}(1999)}]{vandeSand1999}%
  \BibitemOpen
  \bibfield  {author} {\bibinfo {author} {\bibfnamefont {G.}~\bibnamefont
  {van~de Sand}}\ and\ \bibinfo {author} {\bibfnamefont {J.~M.}\ \bibnamefont
  {Rost}},\ }\bibfield  {title} {\bibinfo {title} {Irregular orbits generate
  higher harmonics},\ }\href@noop {} {\bibfield  {journal} {\bibinfo  {journal}
  {Phys. Rev. Lett.}\ }\textbf {\bibinfo {volume} {83}},\ \bibinfo {pages}
  {524} (\bibinfo {year} {1999})}\BibitemShut {NoStop}%
\bibitem [{\citenamefont {Mauger}\ \emph {et~al.}(2009)\citenamefont {Mauger},
  \citenamefont {Chandre},\ and\ \citenamefont {Uzer}}]{Mauger2009}%
  \BibitemOpen
  \bibfield  {author} {\bibinfo {author} {\bibfnamefont {F.}~\bibnamefont
  {Mauger}}, \bibinfo {author} {\bibfnamefont {C.}~\bibnamefont {Chandre}},\
  and\ \bibinfo {author} {\bibfnamefont {T.}~\bibnamefont {Uzer}},\ }\bibfield
  {title} {\bibinfo {title} {Strong field double ionization: The phase space
  perspective},\ }\href@noop {} {\bibfield  {journal} {\bibinfo  {journal}
  {Phys. Rev. Lett.}\ }\textbf {\bibinfo {volume} {102}},\ \bibinfo {pages}
  {173002} (\bibinfo {year} {2009})}\BibitemShut {NoStop}%
\bibitem [{\citenamefont {Mauger}\ \emph {et~al.}(2010)\citenamefont {Mauger},
  \citenamefont {Chandre},\ and\ \citenamefont {Uzer}}]{Mauger2010}%
  \BibitemOpen
  \bibfield  {author} {\bibinfo {author} {\bibfnamefont {F.}~\bibnamefont
  {Mauger}}, \bibinfo {author} {\bibfnamefont {C.}~\bibnamefont {Chandre}},\
  and\ \bibinfo {author} {\bibfnamefont {T.}~\bibnamefont {Uzer}},\ }\bibfield
  {title} {\bibinfo {title} {Recollisions and correlated double ionization with
  circularly polarized light},\ }\href@noop {} {\bibfield  {journal} {\bibinfo
  {journal} {Phys. Rev. Lett.}\ }\textbf {\bibinfo {volume} {105}},\ \bibinfo
  {pages} {083002} (\bibinfo {year} {2010})}\BibitemShut {NoStop}%
\bibitem [{\citenamefont {Becker}\ \emph {et~al.}(2012)\citenamefont {Becker},
  \citenamefont {Liu}, \citenamefont {Ho},\ and\ \citenamefont
  {Eberly}}]{Becker2012}%
  \BibitemOpen
  \bibfield  {author} {\bibinfo {author} {\bibfnamefont {W.}~\bibnamefont
  {Becker}}, \bibinfo {author} {\bibfnamefont {X.}~\bibnamefont {Liu}},
  \bibinfo {author} {\bibfnamefont {P.~J.}\ \bibnamefont {Ho}},\ and\ \bibinfo
  {author} {\bibfnamefont {J.~H.}\ \bibnamefont {Eberly}},\ }\bibfield  {title}
  {\bibinfo {title} {Theories of photoelectron correlation in laser-driven
  multiple atomic ionization},\ }\href@noop {} {\bibfield  {journal} {\bibinfo
  {journal} {Rev. Mod. Phys.}\ }\textbf {\bibinfo {volume} {84}},\ \bibinfo
  {pages} {1011} (\bibinfo {year} {2012})}\BibitemShut {NoStop}%
\bibitem [{\citenamefont {S\'andor}\ \emph {et~al.}(2019)\citenamefont
  {S\'andor}, \citenamefont {Sissay}, \citenamefont {Mauger}, \citenamefont
  {Gordon}, \citenamefont {Gorman}, \citenamefont {Scarborough}, \citenamefont
  {Gaarde}, \citenamefont {Lopata}, \citenamefont {Schafer},\ and\
  \citenamefont {Jones}}]{Sandor2019}%
  \BibitemOpen
  \bibfield  {author} {\bibinfo {author} {\bibfnamefont {P.}~\bibnamefont
  {S\'andor}}, \bibinfo {author} {\bibfnamefont {A.}~\bibnamefont {Sissay}},
  \bibinfo {author} {\bibfnamefont {F.}~\bibnamefont {Mauger}}, \bibinfo
  {author} {\bibfnamefont {M.~W.}\ \bibnamefont {Gordon}}, \bibinfo {author}
  {\bibfnamefont {T.~T.}\ \bibnamefont {Gorman}}, \bibinfo {author}
  {\bibfnamefont {T.~D.}\ \bibnamefont {Scarborough}}, \bibinfo {author}
  {\bibfnamefont {M.~B.}\ \bibnamefont {Gaarde}}, \bibinfo {author}
  {\bibfnamefont {K.}~\bibnamefont {Lopata}}, \bibinfo {author} {\bibfnamefont
  {K.~J.}\ \bibnamefont {Schafer}},\ and\ \bibinfo {author} {\bibfnamefont
  {R.~R.}\ \bibnamefont {Jones}},\ }\bibfield  {title} {\bibinfo {title}
  {Angle-dependent strong-field ionization of halomethanes},\ }\href@noop {}
  {\bibfield  {journal} {\bibinfo  {journal} {J. Chem. Phys.}\ }\textbf
  {\bibinfo {volume} {151}},\ \bibinfo {pages} {194308} (\bibinfo {year}
  {2019})}\BibitemShut {NoStop}%
\bibitem [{\citenamefont {Tuthill}\ \emph {et~al.}(2020)\citenamefont
  {Tuthill}, \citenamefont {Mauger}, \citenamefont {Scarborough}, \citenamefont
  {Jones}, \citenamefont {Gaarde}, \citenamefont {Lopata}, \citenamefont
  {Schafer},\ and\ \citenamefont {DiMauro}}]{Tuthill2020}%
  \BibitemOpen
  \bibfield  {author} {\bibinfo {author} {\bibfnamefont {D.~R.}\ \bibnamefont
  {Tuthill}}, \bibinfo {author} {\bibfnamefont {F.}~\bibnamefont {Mauger}},
  \bibinfo {author} {\bibfnamefont {T.~D.}\ \bibnamefont {Scarborough}},
  \bibinfo {author} {\bibfnamefont {R.~R.}\ \bibnamefont {Jones}}, \bibinfo
  {author} {\bibfnamefont {M.~B.}\ \bibnamefont {Gaarde}}, \bibinfo {author}
  {\bibfnamefont {K.}~\bibnamefont {Lopata}}, \bibinfo {author} {\bibfnamefont
  {K.~J.}\ \bibnamefont {Schafer}},\ and\ \bibinfo {author} {\bibfnamefont
  {L.~F.}\ \bibnamefont {DiMauro}},\ }\bibfield  {title} {\bibinfo {title}
  {Multidimensional molecular high-harmonic spectroscopy: A road map for charge
  migration studies},\ }\href@noop {} {\bibfield  {journal} {\bibinfo
  {journal} {J. Mol. Spectro.}\ }\textbf {\bibinfo {volume} {372}},\ \bibinfo
  {pages} {111353} (\bibinfo {year} {2020})}\BibitemShut {NoStop}%
\bibitem [{\citenamefont {Kohn}\ and\ \citenamefont {Sham}(1965)}]{Kohn1965}%
  \BibitemOpen
  \bibfield  {author} {\bibinfo {author} {\bibfnamefont {W.}~\bibnamefont
  {Kohn}}\ and\ \bibinfo {author} {\bibfnamefont {L.~J.}\ \bibnamefont
  {Sham}},\ }\bibfield  {title} {\bibinfo {title} {Self-consistent equations
  including exchange and correlation effects},\ }\href@noop {} {\bibfield
  {journal} {\bibinfo  {journal} {Phys. Rev.}\ }\textbf {\bibinfo {volume}
  {140}},\ \bibinfo {pages} {A1133} (\bibinfo {year} {1965})}\BibitemShut
  {NoStop}%
\bibitem [{\citenamefont {Eshuis}\ and\ \citenamefont {van
  Voorhis}(2009)}]{Eshuis2009}%
  \BibitemOpen
  \bibfield  {author} {\bibinfo {author} {\bibfnamefont {H.}~\bibnamefont
  {Eshuis}}\ and\ \bibinfo {author} {\bibfnamefont {T.}~\bibnamefont {van
  Voorhis}},\ }\bibfield  {title} {\bibinfo {title} {The influence of initial
  conditions on charge transfer dynamics},\ }\href@noop {} {\bibfield
  {journal} {\bibinfo  {journal} {Phys. Chem. Chem. Phys.}\ }\textbf {\bibinfo
  {volume} {11}},\ \bibinfo {pages} {10293} (\bibinfo {year}
  {2009})}\BibitemShut {NoStop}%
\bibitem [{\citenamefont {Laskar}(1993)}]{Laskar1993}%
  \BibitemOpen
  \bibfield  {author} {\bibinfo {author} {\bibfnamefont {J.}~\bibnamefont
  {Laskar}},\ }\bibfield  {title} {\bibinfo {title} {Frequency analysis for
  multi-dimensional systems. {G}lobal dynamics and diffusion},\ }\href@noop {}
  {\bibfield  {journal} {\bibinfo  {journal} {Physica D}\ }\textbf {\bibinfo
  {volume} {67}},\ \bibinfo {pages} {257} (\bibinfo {year} {1993})}\BibitemShut
  {NoStop}%
\bibitem [{\citenamefont {Laskar}(1999)}]{Laskar1999}%
  \BibitemOpen
  \bibfield  {author} {\bibinfo {author} {\bibfnamefont {J.}~\bibnamefont
  {Laskar}},\ }\bibinfo {title} {Introduction to frequency map analysis, {I}n:
  {H}amiltonian systems with three or more degrees of freedom},\ in\ \href
  {https://doi.org/10.1007/978-94-011-4673-9\_13} {\emph {\bibinfo {booktitle}
  {Hamiltonian Systems with Three or More Degrees of Freedom}}},\ \bibinfo
  {editor} {edited by\ \bibinfo {editor} {\bibfnamefont {C.}~\bibnamefont
  {Sim{\'o}}}}\ (\bibinfo  {publisher} {Springer Netherlands},\ \bibinfo {year}
  {1999})\ p.\ \bibinfo {pages} {134}\BibitemShut {NoStop}%
\bibitem [{\citenamefont {Laskar}\ and\ \citenamefont
  {Robutel}(1993)}]{Laskar1993_2}%
  \BibitemOpen
  \bibfield  {author} {\bibinfo {author} {\bibfnamefont {J.}~\bibnamefont
  {Laskar}}\ and\ \bibinfo {author} {\bibfnamefont {P.}~\bibnamefont
  {Robutel}},\ }\bibfield  {title} {\bibinfo {title} {The chaotic obliquity of
  the planets},\ }\href@noop {} {\bibfield  {journal} {\bibinfo  {journal}
  {Nature}\ }\textbf {\bibinfo {volume} {361}},\ \bibinfo {pages} {608}
  (\bibinfo {year} {1993})}\BibitemShut {NoStop}%
\bibitem [{\citenamefont {Laskar}\ \emph {et~al.}(1993)\citenamefont {Laskar},
  \citenamefont {Joutel},\ and\ \citenamefont {Robutel}}]{Laskar1993_3}%
  \BibitemOpen
  \bibfield  {author} {\bibinfo {author} {\bibfnamefont {J.}~\bibnamefont
  {Laskar}}, \bibinfo {author} {\bibfnamefont {F.}~\bibnamefont {Joutel}},\
  and\ \bibinfo {author} {\bibfnamefont {P.}~\bibnamefont {Robutel}},\
  }\bibfield  {title} {\bibinfo {title} {Stabilization of the {E}arth's
  obliquity by the {M}oon},\ }\href@noop {} {\bibfield  {journal} {\bibinfo
  {journal} {Nature}\ }\textbf {\bibinfo {volume} {361}},\ \bibinfo {pages}
  {615} (\bibinfo {year} {1993})}\BibitemShut {NoStop}%
\bibitem [{\citenamefont {Lopata}\ and\ \citenamefont
  {Govind}(2011)}]{Lopata2011}%
  \BibitemOpen
  \bibfield  {author} {\bibinfo {author} {\bibfnamefont {K.}~\bibnamefont
  {Lopata}}\ and\ \bibinfo {author} {\bibfnamefont {N.}~\bibnamefont
  {Govind}},\ }\bibfield  {title} {\bibinfo {title} {Modeling fast electron
  dynamics with real-time time-dependent density functional theory: Application
  to small molecules and chromophores},\ }\href@noop {} {\bibfield  {journal}
  {\bibinfo  {journal} {J. Chem. Theory Comput.}\ }\textbf {\bibinfo {volume}
  {7}},\ \bibinfo {pages} {1344} (\bibinfo {year} {2011})}\BibitemShut
  {NoStop}%
\bibitem [{\citenamefont {Apr\`a}\ \emph {et~al.}(2020)\citenamefont {Apr\`a},
  \citenamefont {Bylaska}, \citenamefont {de~Jong}, \citenamefont {Govind},
  \citenamefont {Kowalski}, \citenamefont {Straatsma}, \citenamefont {Valiev},
  \citenamefont {van Dam}, \citenamefont {Alexeev}, \citenamefont {Anchell},
  \citenamefont {Anisimov}, \citenamefont {Aquino}, \citenamefont {Atta-Fynn},
  \citenamefont {Autschbach}, \citenamefont {Bauman}, \citenamefont {Becca},
  \citenamefont {Bernholdt}, \citenamefont {Bhaskaran-Nair}, \citenamefont
  {Bogatko}, \citenamefont {Borowski}, \citenamefont {Boschen}, \citenamefont
  {Brabec}, \citenamefont {Bruner}, \citenamefont {Cau\"et}, \citenamefont
  {Chen}, \citenamefont {Chuev}, \citenamefont {Cramer}, \citenamefont {Daily},
  \citenamefont {Deegan}, \citenamefont {Dunning}, \citenamefont {Dupuis},
  \citenamefont {Dyall}, \citenamefont {Fann}, \citenamefont {Fischer},
  \citenamefont {Fonari}, \citenamefont {Fr\"uchtl}, \citenamefont {Gagliardi},
  \citenamefont {Garza}, \citenamefont {Gawande}, \citenamefont {Ghosh},
  \citenamefont {Glaesemann}, \citenamefont {G\"otz}, \citenamefont {Hammond},
  \citenamefont {Helms}, \citenamefont {Hermes}, \citenamefont {Hirao},
  \citenamefont {Hirata}, \citenamefont {Jacquelin}, \citenamefont {Jensen},
  \citenamefont {Johnson}, \citenamefont {J\'onsson}, \citenamefont {Kendall},
  \citenamefont {Klemm}, \citenamefont {Kobayashi}, \citenamefont {Konkov},
  \citenamefont {Krishnamoorthy}, \citenamefont {Krishnan}, \citenamefont
  {Lin}, \citenamefont {Lins}, \citenamefont {Littlefield}, \citenamefont
  {Logsdail}, \citenamefont {Lopata}, \citenamefont {Ma}, \citenamefont
  {Marenich}, \citenamefont {Martin~del Campo}, \citenamefont
  {Mejia-Rodriguez}, \citenamefont {Moore}, \citenamefont {Mullin},
  \citenamefont {Nakajima}, \citenamefont {Nascimento}, \citenamefont
  {Nichols}, \citenamefont {Nichols}, \citenamefont {Nieplocha}, \citenamefont
  {Otero-de-la Roza}, \citenamefont {Palmer}, \citenamefont {Panyala},
  \citenamefont {Pirojsirikul}, \citenamefont {Peng}, \citenamefont {Peverati},
  \citenamefont {Pittner}, \citenamefont {Pollack}, \citenamefont {Richard},
  \citenamefont {Sadayappan}, \citenamefont {Schatz}, \citenamefont {Shelton},
  \citenamefont {Silverstein}, \citenamefont {Smith}, \citenamefont {Soares},
  \citenamefont {Song}, \citenamefont {Swart}, \citenamefont {Taylor},
  \citenamefont {Thomas}, \citenamefont {Tipparaju}, \citenamefont {Truhlar},
  \citenamefont {Tsemekhman}, \citenamefont {Van~Voorhis}, \citenamefont
  {V\'azquez-Mayagoitia}, \citenamefont {Verma}, \citenamefont {Villa},
  \citenamefont {Vishnu}, \citenamefont {Vogiatzis}, \citenamefont {Wang},
  \citenamefont {Weare}, \citenamefont {Williamson}, \citenamefont {Windus},
  \citenamefont {Woli\'nski}, \citenamefont {Wong}, \citenamefont {Wu},
  \citenamefont {Yang}, \citenamefont {Yu}, \citenamefont {Zacharias},
  \citenamefont {Zhang}, \citenamefont {Zhao},\ and\ \citenamefont
  {Harrison}}]{Apra2020}%
  \BibitemOpen
  \bibfield  {author} {\bibinfo {author} {\bibfnamefont {E.}~\bibnamefont
  {Apr\`a}}, \bibinfo {author} {\bibfnamefont {E.~J.}\ \bibnamefont {Bylaska}},
  \bibinfo {author} {\bibfnamefont {W.~A.}\ \bibnamefont {de~Jong}}, \bibinfo
  {author} {\bibfnamefont {N.}~\bibnamefont {Govind}}, \bibinfo {author}
  {\bibfnamefont {K.}~\bibnamefont {Kowalski}}, \bibinfo {author}
  {\bibfnamefont {T.~P.}\ \bibnamefont {Straatsma}}, \bibinfo {author}
  {\bibfnamefont {M.}~\bibnamefont {Valiev}}, \bibinfo {author} {\bibfnamefont
  {H.~J.~J.}\ \bibnamefont {van Dam}}, \bibinfo {author} {\bibfnamefont
  {Y.}~\bibnamefont {Alexeev}}, \bibinfo {author} {\bibfnamefont
  {J.}~\bibnamefont {Anchell}}, \bibinfo {author} {\bibfnamefont
  {V.}~\bibnamefont {Anisimov}}, \bibinfo {author} {\bibfnamefont {F.~W.}\
  \bibnamefont {Aquino}}, \bibinfo {author} {\bibfnamefont {R.}~\bibnamefont
  {Atta-Fynn}}, \bibinfo {author} {\bibfnamefont {J.}~\bibnamefont
  {Autschbach}}, \bibinfo {author} {\bibfnamefont {N.~P.}\ \bibnamefont
  {Bauman}}, \bibinfo {author} {\bibfnamefont {J.~C.}\ \bibnamefont {Becca}},
  \bibinfo {author} {\bibfnamefont {D.~E.}\ \bibnamefont {Bernholdt}}, \bibinfo
  {author} {\bibfnamefont {K.}~\bibnamefont {Bhaskaran-Nair}}, \bibinfo
  {author} {\bibfnamefont {S.}~\bibnamefont {Bogatko}}, \bibinfo {author}
  {\bibfnamefont {P.}~\bibnamefont {Borowski}}, \bibinfo {author}
  {\bibfnamefont {J.}~\bibnamefont {Boschen}}, \bibinfo {author} {\bibfnamefont
  {J.}~\bibnamefont {Brabec}}, \bibinfo {author} {\bibfnamefont
  {A.}~\bibnamefont {Bruner}}, \bibinfo {author} {\bibfnamefont
  {E.}~\bibnamefont {Cau\"et}}, \bibinfo {author} {\bibfnamefont
  {Y.}~\bibnamefont {Chen}}, \bibinfo {author} {\bibfnamefont {G.~N.}\
  \bibnamefont {Chuev}}, \bibinfo {author} {\bibfnamefont {C.~J.}\ \bibnamefont
  {Cramer}}, \bibinfo {author} {\bibfnamefont {J.}~\bibnamefont {Daily}},
  \bibinfo {author} {\bibfnamefont {M.~J.~O.}\ \bibnamefont {Deegan}}, \bibinfo
  {author} {\bibfnamefont {T.~H.}\ \bibnamefont {Dunning}}, \bibinfo {author}
  {\bibfnamefont {M.}~\bibnamefont {Dupuis}}, \bibinfo {author} {\bibfnamefont
  {K.~G.}\ \bibnamefont {Dyall}}, \bibinfo {author} {\bibfnamefont {G.~I.}\
  \bibnamefont {Fann}}, \bibinfo {author} {\bibfnamefont {S.~A.}\ \bibnamefont
  {Fischer}}, \bibinfo {author} {\bibfnamefont {A.}~\bibnamefont {Fonari}},
  \bibinfo {author} {\bibfnamefont {H.}~\bibnamefont {Fr\"uchtl}}, \bibinfo
  {author} {\bibfnamefont {L.}~\bibnamefont {Gagliardi}}, \bibinfo {author}
  {\bibfnamefont {J.}~\bibnamefont {Garza}}, \bibinfo {author} {\bibfnamefont
  {N.}~\bibnamefont {Gawande}}, \bibinfo {author} {\bibfnamefont
  {S.}~\bibnamefont {Ghosh}}, \bibinfo {author} {\bibfnamefont
  {K.}~\bibnamefont {Glaesemann}}, \bibinfo {author} {\bibfnamefont {A.~W.}\
  \bibnamefont {G\"otz}}, \bibinfo {author} {\bibfnamefont {J.}~\bibnamefont
  {Hammond}}, \bibinfo {author} {\bibfnamefont {V.}~\bibnamefont {Helms}},
  \bibinfo {author} {\bibfnamefont {E.~D.}\ \bibnamefont {Hermes}}, \bibinfo
  {author} {\bibfnamefont {K.}~\bibnamefont {Hirao}}, \bibinfo {author}
  {\bibfnamefont {S.}~\bibnamefont {Hirata}}, \bibinfo {author} {\bibfnamefont
  {M.}~\bibnamefont {Jacquelin}}, \bibinfo {author} {\bibfnamefont
  {L.}~\bibnamefont {Jensen}}, \bibinfo {author} {\bibfnamefont {B.~G.}\
  \bibnamefont {Johnson}}, \bibinfo {author} {\bibfnamefont {H.}~\bibnamefont
  {J\'onsson}}, \bibinfo {author} {\bibfnamefont {R.~A.}\ \bibnamefont
  {Kendall}}, \bibinfo {author} {\bibfnamefont {M.}~\bibnamefont {Klemm}},
  \bibinfo {author} {\bibfnamefont {R.}~\bibnamefont {Kobayashi}}, \bibinfo
  {author} {\bibfnamefont {V.}~\bibnamefont {Konkov}}, \bibinfo {author}
  {\bibfnamefont {S.}~\bibnamefont {Krishnamoorthy}}, \bibinfo {author}
  {\bibfnamefont {M.}~\bibnamefont {Krishnan}}, \bibinfo {author}
  {\bibfnamefont {Z.}~\bibnamefont {Lin}}, \bibinfo {author} {\bibfnamefont
  {R.~D.}\ \bibnamefont {Lins}}, \bibinfo {author} {\bibfnamefont {R.~J.}\
  \bibnamefont {Littlefield}}, \bibinfo {author} {\bibfnamefont {A.~J.}\
  \bibnamefont {Logsdail}}, \bibinfo {author} {\bibfnamefont {K.}~\bibnamefont
  {Lopata}}, \bibinfo {author} {\bibfnamefont {W.}~\bibnamefont {Ma}}, \bibinfo
  {author} {\bibfnamefont {A.~V.}\ \bibnamefont {Marenich}}, \bibinfo {author}
  {\bibfnamefont {J.}~\bibnamefont {Martin~del Campo}}, \bibinfo {author}
  {\bibfnamefont {D.}~\bibnamefont {Mejia-Rodriguez}}, \bibinfo {author}
  {\bibfnamefont {J.~E.}\ \bibnamefont {Moore}}, \bibinfo {author}
  {\bibfnamefont {J.~M.}\ \bibnamefont {Mullin}}, \bibinfo {author}
  {\bibfnamefont {T.}~\bibnamefont {Nakajima}}, \bibinfo {author}
  {\bibfnamefont {D.~R.}\ \bibnamefont {Nascimento}}, \bibinfo {author}
  {\bibfnamefont {J.~A.}\ \bibnamefont {Nichols}}, \bibinfo {author}
  {\bibfnamefont {P.~J.}\ \bibnamefont {Nichols}}, \bibinfo {author}
  {\bibfnamefont {J.}~\bibnamefont {Nieplocha}}, \bibinfo {author}
  {\bibfnamefont {A.}~\bibnamefont {Otero-de-la Roza}}, \bibinfo {author}
  {\bibfnamefont {B.}~\bibnamefont {Palmer}}, \bibinfo {author} {\bibfnamefont
  {A.}~\bibnamefont {Panyala}}, \bibinfo {author} {\bibfnamefont
  {T.}~\bibnamefont {Pirojsirikul}}, \bibinfo {author} {\bibfnamefont
  {B.}~\bibnamefont {Peng}}, \bibinfo {author} {\bibfnamefont {R.}~\bibnamefont
  {Peverati}}, \bibinfo {author} {\bibfnamefont {J.}~\bibnamefont {Pittner}},
  \bibinfo {author} {\bibfnamefont {L.}~\bibnamefont {Pollack}}, \bibinfo
  {author} {\bibfnamefont {R.~M.}\ \bibnamefont {Richard}}, \bibinfo {author}
  {\bibfnamefont {P.}~\bibnamefont {Sadayappan}}, \bibinfo {author}
  {\bibfnamefont {G.~C.}\ \bibnamefont {Schatz}}, \bibinfo {author}
  {\bibfnamefont {W.~A.}\ \bibnamefont {Shelton}}, \bibinfo {author}
  {\bibfnamefont {D.~W.}\ \bibnamefont {Silverstein}}, \bibinfo {author}
  {\bibfnamefont {D.~M.~A.}\ \bibnamefont {Smith}}, \bibinfo {author}
  {\bibfnamefont {T.~A.}\ \bibnamefont {Soares}}, \bibinfo {author}
  {\bibfnamefont {D.}~\bibnamefont {Song}}, \bibinfo {author} {\bibfnamefont
  {M.}~\bibnamefont {Swart}}, \bibinfo {author} {\bibfnamefont {H.~L.}\
  \bibnamefont {Taylor}}, \bibinfo {author} {\bibfnamefont {G.~S.}\
  \bibnamefont {Thomas}}, \bibinfo {author} {\bibfnamefont {V.}~\bibnamefont
  {Tipparaju}}, \bibinfo {author} {\bibfnamefont {D.~G.}\ \bibnamefont
  {Truhlar}}, \bibinfo {author} {\bibfnamefont {K.}~\bibnamefont {Tsemekhman}},
  \bibinfo {author} {\bibfnamefont {T.}~\bibnamefont {Van~Voorhis}}, \bibinfo
  {author} {\bibfnamefont {A.}~\bibnamefont {V\'azquez-Mayagoitia}}, \bibinfo
  {author} {\bibfnamefont {P.}~\bibnamefont {Verma}}, \bibinfo {author}
  {\bibfnamefont {O.}~\bibnamefont {Villa}}, \bibinfo {author} {\bibfnamefont
  {A.}~\bibnamefont {Vishnu}}, \bibinfo {author} {\bibfnamefont {K.~D.}\
  \bibnamefont {Vogiatzis}}, \bibinfo {author} {\bibfnamefont {D.}~\bibnamefont
  {Wang}}, \bibinfo {author} {\bibfnamefont {J.~H.}\ \bibnamefont {Weare}},
  \bibinfo {author} {\bibfnamefont {M.~J.}\ \bibnamefont {Williamson}},
  \bibinfo {author} {\bibfnamefont {T.~L.}\ \bibnamefont {Windus}}, \bibinfo
  {author} {\bibfnamefont {K.}~\bibnamefont {Woli\'nski}}, \bibinfo {author}
  {\bibfnamefont {A.~T.}\ \bibnamefont {Wong}}, \bibinfo {author}
  {\bibfnamefont {Q.}~\bibnamefont {Wu}}, \bibinfo {author} {\bibfnamefont
  {C.}~\bibnamefont {Yang}}, \bibinfo {author} {\bibfnamefont {Q.}~\bibnamefont
  {Yu}}, \bibinfo {author} {\bibfnamefont {M.}~\bibnamefont {Zacharias}},
  \bibinfo {author} {\bibfnamefont {Z.}~\bibnamefont {Zhang}}, \bibinfo
  {author} {\bibfnamefont {Y.}~\bibnamefont {Zhao}},\ and\ \bibinfo {author}
  {\bibfnamefont {R.~J.}\ \bibnamefont {Harrison}},\ }\bibfield  {title}
  {\bibinfo {title} {{NWC}hem: Past, present, and future},\ }\href@noop {}
  {\bibfield  {journal} {\bibinfo  {journal} {J. Chem. Phys.}\ }\textbf
  {\bibinfo {volume} {152}},\ \bibinfo {pages} {184102} (\bibinfo {year}
  {2020})}\BibitemShut {NoStop}%
\bibitem [{\citenamefont {Acebr\'on}\ \emph {et~al.}(2005)\citenamefont
  {Acebr\'on}, \citenamefont {Bonilla}, \citenamefont {P\'erez~Vicente},
  \citenamefont {Ritort},\ and\ \citenamefont {Spigler}}]{Acebron2005}%
  \BibitemOpen
  \bibfield  {author} {\bibinfo {author} {\bibfnamefont {J.~A.}\ \bibnamefont
  {Acebr\'on}}, \bibinfo {author} {\bibfnamefont {L.~L.}\ \bibnamefont
  {Bonilla}}, \bibinfo {author} {\bibfnamefont {C.~J.}\ \bibnamefont
  {P\'erez~Vicente}}, \bibinfo {author} {\bibfnamefont {F.}~\bibnamefont
  {Ritort}},\ and\ \bibinfo {author} {\bibfnamefont {R.}~\bibnamefont
  {Spigler}},\ }\bibfield  {title} {\bibinfo {title} {The {K}uramoto model: A
  simple paradigm for synchronization phenomena},\ }\href@noop {} {\bibfield
  {journal} {\bibinfo  {journal} {Rev. Mod. Phys.}\ }\textbf {\bibinfo {volume}
  {77}},\ \bibinfo {pages} {137} (\bibinfo {year} {2005})}\BibitemShut
  {NoStop}%
\bibitem [{\citenamefont {Feldhaus}\ \emph {et~al.}(2005)\citenamefont
  {Feldhaus}, \citenamefont {Arthur},\ and\ \citenamefont
  {Hastings}}]{Feldhaus2005}%
  \BibitemOpen
  \bibfield  {author} {\bibinfo {author} {\bibfnamefont {J.}~\bibnamefont
  {Feldhaus}}, \bibinfo {author} {\bibfnamefont {J.}~\bibnamefont {Arthur}},\
  and\ \bibinfo {author} {\bibfnamefont {J.}~\bibnamefont {Hastings}},\
  }\bibfield  {title} {\bibinfo {title} {X-ray free-electron lasers},\
  }\href@noop {} {\bibfield  {journal} {\bibinfo  {journal} {J. Phys. B: At.
  Mol. Opt. Phys.}\ }\textbf {\bibinfo {volume} {38}},\ \bibinfo {pages} {S799}
  (\bibinfo {year} {2005})}\BibitemShut {NoStop}%
\bibitem [{\citenamefont {Pikovsky}\ and\ \citenamefont
  {Rosenblum}(2007)}]{Pikovsky2007}%
  \BibitemOpen
  \bibfield  {author} {\bibinfo {author} {\bibfnamefont {A.}~\bibnamefont
  {Pikovsky}}\ and\ \bibinfo {author} {\bibfnamefont {M.}~\bibnamefont
  {Rosenblum}},\ }\bibfield  {title} {\bibinfo {title} {Synchronization},\
  }\href@noop {} {\bibfield  {journal} {\bibinfo  {journal} {Scholarpedia}\
  }\textbf {\bibinfo {volume} {2}},\ \bibinfo {pages} {1459} (\bibinfo {year}
  {2007})}\BibitemShut {NoStop}%
\bibitem [{\citenamefont {Javanainen}\ \emph {et~al.}(1988)\citenamefont
  {Javanainen}, \citenamefont {Eberly},\ and\ \citenamefont
  {Su}}]{Javanainen1988}%
  \BibitemOpen
  \bibfield  {author} {\bibinfo {author} {\bibfnamefont {J.}~\bibnamefont
  {Javanainen}}, \bibinfo {author} {\bibfnamefont {J.~H.}\ \bibnamefont
  {Eberly}},\ and\ \bibinfo {author} {\bibfnamefont {Q.}~\bibnamefont {Su}},\
  }\bibfield  {title} {\bibinfo {title} {Numerical simulations of multiphoton
  ionization and above-threshold electron spectra},\ }\href@noop {} {\bibfield
  {journal} {\bibinfo  {journal} {Phys. Rev. A}\ }\textbf {\bibinfo {volume}
  {38}},\ \bibinfo {pages} {3430} (\bibinfo {year} {1988})}\BibitemShut
  {NoStop}%
\bibitem [{\citenamefont {Helbig}\ \emph {et~al.}(2011)\citenamefont {Helbig},
  \citenamefont {Fuks}, \citenamefont {Casula}, \citenamefont {Verstraete},
  \citenamefont {Marques}, \citenamefont {Tokatly},\ and\ \citenamefont
  {Rubio}}]{Helbig2011}%
  \BibitemOpen
  \bibfield  {author} {\bibinfo {author} {\bibfnamefont {N.}~\bibnamefont
  {Helbig}}, \bibinfo {author} {\bibfnamefont {J.~I.}\ \bibnamefont {Fuks}},
  \bibinfo {author} {\bibfnamefont {M.}~\bibnamefont {Casula}}, \bibinfo
  {author} {\bibfnamefont {M.~J.}\ \bibnamefont {Verstraete}}, \bibinfo
  {author} {\bibfnamefont {M.~A.~L.}\ \bibnamefont {Marques}}, \bibinfo
  {author} {\bibfnamefont {I.~V.}\ \bibnamefont {Tokatly}},\ and\ \bibinfo
  {author} {\bibfnamefont {A.}~\bibnamefont {Rubio}},\ }\bibfield  {title}
  {\bibinfo {title} {Density functional theory beyond the linear regime:
  Validating an adiabatic local density approximation},\ }\href@noop {}
  {\bibfield  {journal} {\bibinfo  {journal} {Phys. Rev. A}\ }\textbf {\bibinfo
  {volume} {83}},\ \bibinfo {pages} {032503} (\bibinfo {year}
  {2011})}\BibitemShut {NoStop}%
\bibitem [{\citenamefont {Legrand}\ \emph {et~al.}(2002)\citenamefont
  {Legrand}, \citenamefont {Suraud},\ and\ \citenamefont
  {Reinhard}}]{Legrand2002}%
  \BibitemOpen
  \bibfield  {author} {\bibinfo {author} {\bibfnamefont {C.}~\bibnamefont
  {Legrand}}, \bibinfo {author} {\bibfnamefont {E.}~\bibnamefont {Suraud}},\
  and\ \bibinfo {author} {\bibfnamefont {P.-G.}\ \bibnamefont {Reinhard}},\
  }\bibfield  {title} {\bibinfo {title} {Comparison of
  self-interaction-corrections for metal clusters},\ }\href@noop {} {\bibfield
  {journal} {\bibinfo  {journal} {J. Phys. B: At. Mol. Opt. Phys.}\ }\textbf
  {\bibinfo {volume} {35}},\ \bibinfo {pages} {1115} (\bibinfo {year}
  {2002})}\BibitemShut {NoStop}%
\end{thebibliography}

%

\end{document}